\documentclass[aps,prx,twocolumn,superscriptaddress,longbibliography]{revtex4-1}
\usepackage{graphicx}
\usepackage{latexsym}
\usepackage{amssymb}
\usepackage{amsmath}
\usepackage{amsfonts}
\usepackage{upgreek}
\usepackage{bm}
\usepackage{multirow}
\usepackage{soul}
\usepackage{enumitem}
\usepackage{color}
\usepackage{youngtab}
\usepackage{pict2e}
\makeatletter
\newcommand{\crossout}[1]{%
  \begingroup
  \settowidth{\dimen@}{#1}%
  \setlength{\unitlength}{0.05\dimen@}%
  \settoheight{\dimen@}{#1}%
  \count@=\dimen@
  \divide\count@ by \unitlength
  \begin{picture}(0,0)
  \put(0,0){\line(20,\count@){20}}
  \put(0,\count@){\line(20,-\count@){20}}
  \end{picture}%
  #1%
  \endgroup}
\makeatother
\usepackage{hyperref}
\hypersetup{
colorlinks = true,
linkcolor = [rgb]{0.70,0.13,0.13},
citecolor = [rgb]{0.13,0.55,0.13},
urlcolor = [rgb]{0.25,0.41,0.88}}
\newcommand{\bra}[1]{\langle #1|}
\newcommand{\ket}[1]{|#1 \rangle}
\newcommand{\dd}{\mathrm{d}}
\newcommand{\ii}{\mathrm{i}}
\newcommand{\SU}{\mathrm{SU}}
\newcommand{\U}{\mathrm{U}}
\renewcommand{\O}{\mathrm{O}}
\newcommand{\SO}{\mathrm{SO}}

\newcommand{\dsZ}{\mathbb{Z}}

\newcommand{\scK}{{\mathcal{K}}}
\newcommand{\scL}{{\mathcal{L}}}
\newcommand{\scO}{{\mathcal{O}}}
\newcommand{\scT}{{\mathcal{T}}}
\newcommand{\sfid}{{\mathsf{1}}}
\newcommand{\sfe}{{\mathsf{e}}}
\newcommand{\sfm}{{\mathsf{m}}}
\DeclareSymbolFont{sfletters}{OML}{cmbrm}{m}{it}
\DeclareMathSymbol{\sfeps}{\mathord}{sfletters}{"22}
\newcommand{\Tr}{\operatorname{Tr}}

\renewcommand{\Re}{\operatorname{Re}}
\renewcommand{\Im}{\operatorname{Im}}
\newcommand{\vect}[1]{{\bm{#1}}}

\newcommand{\mat}[1]{\left[\begin{matrix}#1\end{matrix}\right]}
\newcommand{\smat}[1]{\left[\begin{smallmatrix}#1\end{smallmatrix}\right]}

\newcommand{\eqnref}[1]{Eq.\,\eqref{#1}}
\newcommand{\figref}[1]{Fig.\,\ref{#1}}
\newcommand{\tabref}[1]{Tab.\,\ref{#1}}

\begin{document}

\title{Superconductivity from Valley Fluctuations and Approximate SO(4) Symmetry in a Weak Coupling Theory of Twisted Bilayer Graphene}
\author{Yi-Zhuang You}

\author{Ashvin Vishwanath}
\affiliation{Department of Physics, Harvard University, Cambridge, MA 02138, USA}

\date{\today}
\begin{abstract}
The recent discovery of the Mott insulating and superconducting phases in twisted bilayer graphene has generated tremendous research interest.  Here, we develop a weak coupling approach to the superconductivity in twisted bilayer graphene, starting from the Fermi liquid regime. A key observation is that near half filling, the fermiology consists of well nested Fermi pockets derived from opposite valleys, leading to enhanced valley fluctuation, which in turn can mediate superconductivity. This scenario is studied within the random phase approximation. We find that inter-valley electron pairing with either chiral ($d+\ii d$ mixed with $p-\ii p$) or helical form factor is the dominant instability. An  approximate SO(4) spin-valley symmetry  implies a near degeneracy of spin-singlet and triplet pairing. On increasing interactions, commensurate inter-valley coherence wave (IVCW) order can arise, with simultaneous condensation at the three $M$ points in the Brillouin Zone, and a $2\times2$ pattern in real space. In simple treatments though, this  leads to a full gap at fillings $\pm (1/2+1/8)$, slightly away from half-filling .  The selection of spin-singlet or spin triplet orders, both for the IVCW and the superconductor, arise from SO(4) symmetry breaking  terms. Mott insulators derived from phase fluctuating superconductors are also discussed, which exhibit both symmetry protected and intrinsic topological orders. 

\end{abstract}
\maketitle
\section{Introduction}

There has been considerable interest in studying artificial lattices induced by a long wavelength Moir\'e potential in graphene and related materials. These experiments have  recently gathered momentum with the observation of superconductivity and correlated Mott insulators in bilayer graphene twisted to a particular ``magic angle''. The Moir\'e superlattice induced in bilayer graphene twisted by a small angle leads to isolated bands near charge neutrality, whose bandwidth can be tuned by twist angle \cite{Neto2007, STMNatPhy, Magaud2010, Bistritzer2011,  Mele2011, CastroNeto12,STMPRL, Crommie2015, Kim2017, PabloPRL, LeRoy2018, Ensslin2018}. On approaching certain magic angles, the largest being $\sim 1.1^\circ$, the bandwidth is significantly reduced allowing for correlation physics to take hold. Indeed, recent studies on twisted bilayer graphene (tBLG) near the magic angle have revealed the presence of Mott insulators\cite{Cao2018} at fractional filling of the bands, as well as  superconductivity\cite{Cao2018a} in close proximity  to some of the Mott insulators. While Mott physics has also been observed in a  different Moir\'e superlattice system, induced by a boron nitride substrate on ABC trilayer graphene \cite{Wang2018}, here we will focus on the tBLG system, which has already generated a significant amount of theoretical interest \cite{Xu2018,Juricic2018,Volovik2018,Po2018,Fu2018,Kivelson2018,Baskaran2018, Phillips2018, Das2018,Irkhin2018Dirac,Ma2018,Scalettar2018,Yang2018,Zhang2018Low,Zhu2018,Law2018,Vafek2018,Rademaker2018}. 

The band structure of tBLG at small twist angles can be understood from a  continuum model \cite{Neto2007,Bistritzer2011,CastroNeto12} that couples the  Dirac points in the individual graphene layers via the interlayer tunneling. Due to the small twist angles involved, there is a separation of scales between the atomic lattice and the Moir\'e superlattice which implies that commensuration effects can be neglected  \cite{CastroNeto12}. The opposite Dirac points in each layer are  then essentially decoupled, leading to a valley quantum number $n_v =\pm 1$ for each electron ($n_v=+1$ for $K$ valley and $-1$ for $K'$ valley), which is reversed under time reversal symmetry (as valleys are exchanged).  Including both spin and valley degrees of freedom it takes 8 electrons (per Moir\'e unit cell) to completely fill the nearly flat bands that appear near neutrality.  The additional factor of two in the filling  appears due to band contacts present at neutrality and protected by symmetry. Charge neutrality then corresponds to four filled and four empty bands, which meet at Dirac cones at the $K$ points of the Moir\'e Brillouin zone (MBZ). Measuring the electron charge density $n$ from neutrality, the fully filled and fully empty bands occur at $\pm n_s$ ($\sim 2.7 \times 10^{12}{\rm cm}^{-2}$ for magic angle tBLG). In Ref.\,\onlinecite{Cao2018,Cao2018a}, an insulating state was also observed at $f=n/n_s = \mp 1/2$, i.e. at half filling both below and above neutrality (hence the term Mott insulator), where there were two (six) electrons per Moir\'e unit cell.   Furthermore, superconductivity was observed around the $f = -1/2$ Mott insulator, i.e.~around the Mott insulator on the hole doped side of neutrality. 

Although interactions and the band width are both estimated to be comparable in magic angle tBLG, here we consider approaching the problem from the weak coupling limit, i.e.~we imagine moving slightly away from the magic angle, which is motivated by the following considerations.  First, although the energy scale of the bandwidth $W$ \cite{STMNatPhy} and interactions $U$ \cite{Cao2018} were estimated to be of order $10\sim20\text{meV}$, the Mott gap  observed in transport experiments is much smaller $\sim0.4\text{meV}$, and could be closed with an in-plane Zeeman field of  roughly the same strength.   Therefore, the system is  not deep in the Mott regime, where the Mott gap would be of the same order as $U$. Next, doping the Mott insulator towards neutrality results very quickly in a metal with a big Fermi surface, where superconductivity is also observed. This regime could be approached from weak coupling. On the other hand, the side facing the band insulator (i.e. hole doping the $f=-1/2$ Mott insulator or electron doping the $f=+1/2$ Mott insulator) behaves like a `doped' Mott insulator, with both Hall conductivity and quantum oscillations pointing to a small Fermi surface composed of just the doped carries. 

Finally, in both iron-pnictides\cite{Kuroki2008Unconventional,Graser2009Near,Maier2011} and overdoped cuprates\cite{Scalapino1986d-wave,Scalapino1995,Scalapino2012}, weak coupling approaches have been relatively successful at least in predicting the gap symmetry. However, both these calculations relied on band structures with some degree of nesting. Does the fermiology of tBLG support such a nesting driven scenario? Interestingly, on moving slightly away from the magic angle, multiple band structure calculations \cite{CastroNeto12,PabloPRL,ShiangPRB,Cao2018}  for small angle tBLG bands reveal a relatively strong nesting feature in the vicinity of half filling, albeit at  wave-vectors not simply related to the filling. Such nesting is not expected in a single orbital model on the triangular lattice, but appears here quite generally from having opposite valleys  that give rise to a pair of Fermi surfaces related by  time reversal symmetry , each of which is constrained by the microscopic symmetries $C_3$, $M_y$ and $C_2\scT$ as defined in Ref.\,\onlinecite{Po2018}. Within a random phase approximation (RPA), we show that nesting-enhanced valley fluctuations give rise to an inter-valley pairing in the ``$d/p$-wave'' channel ($d$-wave and $p$-wave are generally mixed under $C_3$ symmetry). An important ingredient is the presence of an approximate $\SO(4)$ symmetry. Although four component electrons (spin and valley) might suggest an $\SU(4)$ symmetry, this is strongly broken by the valley-dependent band structure. Instead, we obtain separate spin $\SU(2)$ rotation symmetries for the two valleys $\SU(2)_K\times \SU(2)_{K'}\sim\SO(4)$ with interactions that only depend on the slowly varying part of the electron density. This symmetry ensures a degeneracy of the spin singlet and triplet inter-valley pairing (with valley indices adjusted to ensure the antisymmetry of the pair wave function).  Further weak symmetry breaking terms are expected to split this degeneracy, the experimentally reported Pauli limiting behavior  \cite{Cao2018a} suggests a spin singlet superconductor. This would require invoking a weak anti-Hunds coupling, leading to a inter-valley spin-singlet superconductor with chiral $d+\ii d$ and $p-\ii p$ mixed pairing, while the more conventional Hunds coupling would favor a spin triplet superconductor with chiral $d+\ii d$ / $p-\ii p$ pairing.\cite{Xu2018} Note, in this setting, there is no symmetry distinction between $d+\ii d$ and $p-\ii p$ pairing. 
However, depending on their relative strengths, a topological phase transition occurs characterized by different quantized thermal Hall conductivities (chiral central charge $c=4$ vs $c=-2$). At strong coupling, or with explicit rotation symmetry breaking, nematic superconductivity with two or four nodes may also be stabilized. 

\begin{figure}[htbp]
\begin{center}
\includegraphics[width=0.7\linewidth]{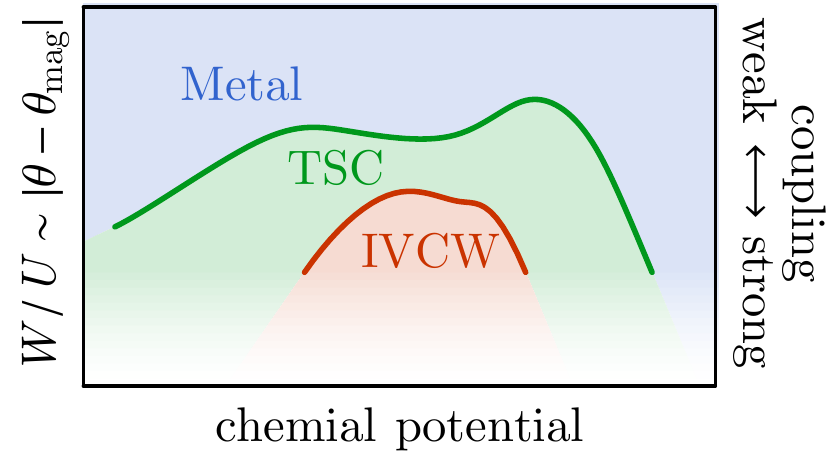}
\caption{Schematic phase diagram in the vicinity of $f=-1/2$, which is obtained by self-consistent mean-field calculation according to \eqnref{eq:HMF} in the  low temperature limit (details will be discussed later). TSC: topological superconductor, IVCW: inter-valley coherence wave. The strong coupling regime (closer to the magic angle) is not captured by this approach. We will mainly focus on the weak coupling regime in this work. The superconductivity is slightly stronger on approaching the van-Hove singularity which is on the electron doped side (neutrality is on the right).}
\label{fig:phases}
\end{center}
\end{figure}

Our general picture is illustrated the phase diagram in \figref{fig:phases}, in the vicinity of $f=-1/2$, which is obtained based on a mean-field model \eqnref{eq:HMF} to be discussed in details later. Tuning the twist angle $\theta$ towards the magic angle $\theta_\text{mag}$ effectively decreases the ratio $W/U$ between the band width $W$ and the interaction $U$, which pushes the system towards strong coupling. Superconductivity will first emerge in the weak coupling regime. At stronger coupling, a simple nesting based picture predicts a inter-valley coherence wave order, with ordering wave vector at the three $M$ points of the MBZ, although a full gap is opened only at filling $f=-(1/2+1/8)$ or at 25\% hole doping. A gap at half filling can open if interactions also modify the electronic dispersion, but this is outside the scope of the present treatment. The RPA approach does not apply to the strong coupling regime (as indicated in the phase diagram by the fading-out color), but we will also comment on alternative approaches that can tackle the strong correlation physics. We should also keep in mind that apart from the ratio $W/U$, the twist angle $\theta$ also influence the band structure especially when $\theta$ gets close to the magic angle $\theta_\text{mag}$. Since the band structure becomes very sensitive to all kinds of perturbations near the magic angle, it is hard to draw universal features right at the magic angle. Thus we will stay a little bit away from the (first) magic angle $\theta_\text{mag}$ by considering $1.2\theta_\text{mag}\lesssim\theta\lesssim2\theta_\text{mag}$, which can provide us a relatively robust band structure and also place us closer into the weak coupling region in the phase diagram \figref{fig:phases}.

{\em Connection to other work:} Given the volume of recent theoretical output  we have to restrict our comments to a few selected references that are closest to this work. Ref.  \cite{Xu2018} starts with an SU(4) Mott insulator, and predicted a topological  superconductor on doping the Mott insulator. Our conclusions are  similar, although we have an SO(4) (rather than SU(4)) symmetry, and we adopt a weak coupling approach which avoids  conflict with localizing electrons  in the narrow bands of tBLG \cite{Po2018}. As in reference \cite{Po2018} we favor a spin-singlet inter-valley ordering, albeit at a finite wave vector, and inter-valley fluctuations drive pairing of a spin-singlet superconductor. Finally, adding strong  SO(4) symmetry breaking terms to our model reproduces the $s$-wave pairing in Ref. \cite{Kivelson2018}.  Although \cite{Yang2018,Scalettar2018,Rademaker2018}  also predicts topological superconductor from weak coupling/quantum Monte Carlo, their models differ significantly from ours. Our proposed pairing mechanism based on the fluctuation of incipient order is similar to Ref. \cite{Juricic2018}, while we identify the leading incipient order to be the valley fluctuation, which differs from the spin fluctuation in \cite{Juricic2018}.

This paper is organized as follows. We start by proposing an effective model for tBLG, deriving the low-energy band structure in Sec.\,\ref{sec:band} and formulating the symmetry-allowed generic interaction in Sec.\,\ref{sec:interaction}. We then analyze the instabilities in all fermion-bilinear channels within the RPA approach in Sec.\,\ref{sec:RPA} and find a leading instability in the inter-valley coherence channel. We study valley fluctuation mediated pairing in Sec.\,\ref{sec:TSC} and identify the dominant superconducting order parameter. We sketch two descriptions for the insulating phase adjacent to the superconducting phase: a Slater insulator with inter-valley coherence wave order in Sec.\,\ref{sec:IVCW} and a topologically ordered Mott insulator obtained by projecting out charge fluctuations of the superconductor in Sec.\,\ref{sec:Mott}. Finally, we study the $\SO(4)$ symmetry breaking effects in Sec.\,\ref{sec:breaking} and close with a discussion in Sec.\,\ref{sec:summary}.

\section{Results}
\subsection{Band Structure and Fermi Surface Nesting}\label{sec:band}

\begin{figure}[htbp]
\begin{center}
\includegraphics[width=0.95\linewidth]{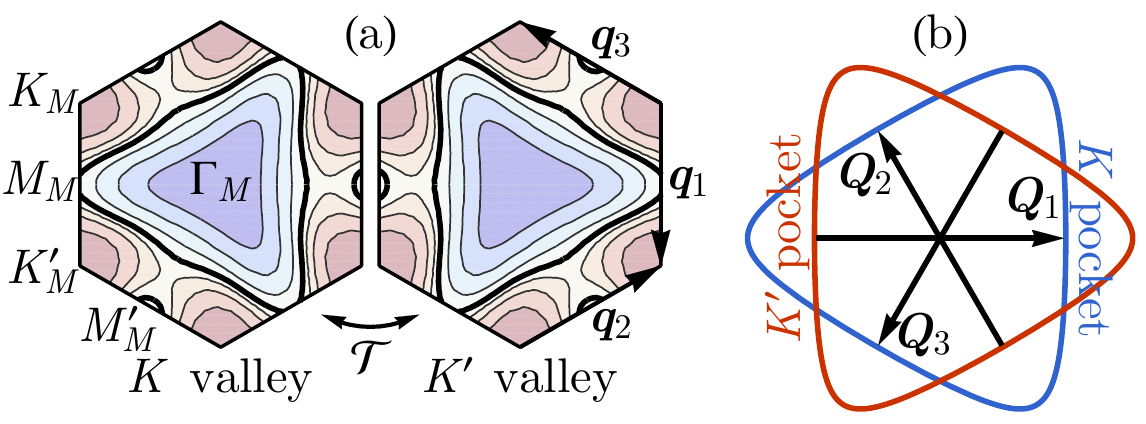}
\caption{(a) Equal-filling contours from the band bottom to the charge neutrality for both valleys in the Moir\'e Brillouin zone. The $-1/2$ filling Fermi surface is traced out by thick lines. (b) The Fermi pockets around $-1/2$ filling are modeled as the triangular shaped Fermi surface in the single-band model. The $K$ and $K'$ pockets are almost nested along three nesting vectors $\vect{Q}_{1,2,3}$.}
\label{fig:pockets}
\end{center}
\end{figure}

We first formulate an effective Hamiltonian that describes the electrons in the Moir\'e band near the Fermi surface. Our starting point is the continuum model of the tBLG proposed in Ref.\,\onlinecite{Bistritzer2011,CastroNeto12}, which first focuses on the band structure around one valley (say the $K$ valley)
\begin{eqnarray}
\label{eq:Hcon}
H_0 &=& H_K+H_{K'}\\ \nonumber
H_K &= & \sum_{\vect{k},l}c_{K_l\vect{k}}^\dagger h_{l\vect{k}} c_{K_l\vect{k}}+\sum_{\vect{k},a}c_{K_-\vect{k}}^\dagger T_{\vect{q}_a} c_{K_+\vect{k}+\vect{q}_a}+\text{h.c.},
\end{eqnarray}
where $c_{K_l\vect{k}}$ denotes the $K$ valley electron originated from the layer $l$ (with $l=\pm1$ labeling the top and the bottom layers respectively). $h_{l\vect{k}}=v_F(\vect{k}-\vect{K}_l)\cdot \vect{\sigma}_{l}$ captures the Dirac dispersion of the electron near the $K_l$ valley, where $\vect{K}_l=R_{\varphi_l}\vect{K}=e^{-\ii \varphi_l\sigma^2}\vect{K}$ is rotated from the monolayer $K$ point $\vect{K}=(4\pi/(3\sqrt{3}),0)$ by an angle $\varphi_l=l\theta/2$ determined by the twist angle $\theta$, and accordingly $\vect{\sigma}_l=e^{-\ii \varphi_l \sigma^3/2}\vect{\sigma}e^{\ii \varphi_l \sigma^3/2}$ is also rotated from the standard Pauli matrices $\vect{\sigma}=(\sigma^1,\sigma^2)$ by the same angle. $T_{\vect{q}_a}=w_0+w_1 (\vect{q}_a\times\vect{\sigma})\cdot\hat{\vect{z}}+\ii w_3 \sigma^3$ describes the interlayer tunneling to the lowest-order of the momentum transfers, as specified by $\vect{q}_1=\vect{K}_{-}-\vect{K}_{+}$, $\vect{q}_2=R_{2\pi/3}\vect{q}_1$ and $\vect{q}_3=R_{-2\pi/3}\vect{q}_1$ in \figref{fig:pockets}(a).  In general, $T_{\vect{q}_a}$ depends on three real parameters $w_0$, $w_1$ and $w_3$ (a typical setting is $w_0\approx w_1 |\vect{q}_a|\gg w_3$).\cite{Neto2007,Bistritzer2011,CastroNeto12} Such a generic form of $T_{\vect{q}_a}$ can be pinned down by symmetry arguments given in Ref.\,\onlinecite{Po2018}. The Hamiltonian $H_{K'}$ around the $K'$ valley can be obtained from $H_{K}$ simply by a time-reversal operation $\scT:c_{K_l\vect{k}}\to \scK c_{K'_l,-\vect{k}}$ (with $\scK$ being the complex conjugation operator). Putting together, $H_0=H_K+H_{K'}$ provides a full description of the low-energy electronic band structure of the tBLG in the continuum limit.

\begin{figure}[htbp]
\begin{center}
\includegraphics[width=0.78\linewidth]{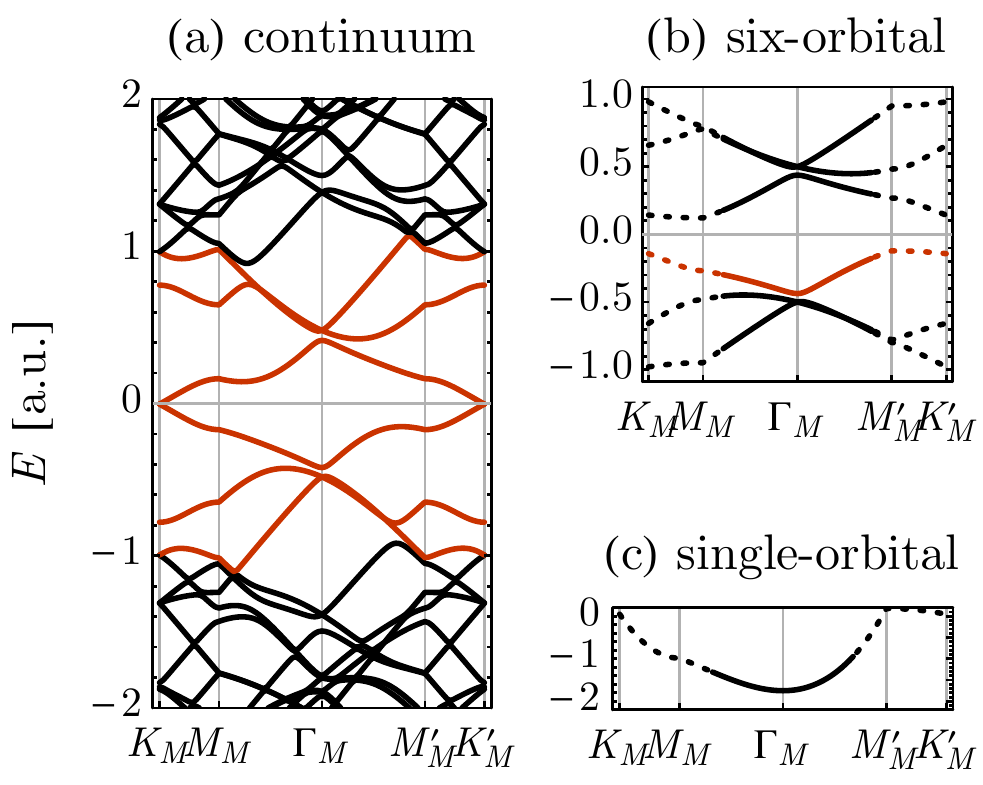}
\caption{Reducing the band structure from (a) the continuum model to (b) the six-orbital model and finally to (c) the single-orbital model. The energy unit is chosen such that $v_F|\vect{q}_a|=1$. The continuum model parameters are taken to be $w_0=0.275$ and $w_1|\vect{q}_a|=-0.3$ for illustration. Each latter model targets the band(s) high-lighted (in red) in the previous model. The reduced models (b,c) are only valid around the Moir\'e $\Gamma_M$ point.}
\label{fig:bands}
\end{center}
\end{figure}

By diagonalizing the Hamiltonian $H_K$ (with an appropriate momentum cutoff), we obtain the single-particle band structure as shown in \figref{fig:bands}(a). The bands are defined in the Moir\'e Brillouin zone (MBZ), as depicted in \figref{fig:pockets}(a) with high symmetry points labeled. The $K_{+}$ and $K_{-}$ valleys from either layers rest on the Moir\'e $K_M$ and $K'_M$ point respectively. We focus on the middle band around the charge neutrality, which will become flat as the twist angle $\theta$ approaches to the magic angles $\theta_\text{mag}$. A prominent feature of this band is that its energy contours (Fermi surfaces) around the $-1/2$ filling typically take triangular shapes around the $\Gamma_M$ point in the MBZ, as shown in \figref{fig:pockets}(a), which was observed in several band theory calculations for small twist angles.\cite{CastroNeto12,PabloPRL,ShiangPRB,Cao2018} The triangular distortion of the Fermi surface is generic on symmetry ground, as it is the lowest order (in terms of angular momentum) distortion that is consistent with all the valley-preserving lattice symmetries $C_6\scT$ and $M_y$.\cite{Po2018,Zou2018Ba} Indeed it is a rather robust feature for a range of twist angles $\theta\gtrsim 1.2\theta_\text{mag}$ and is also stable against perturbations like lattice relaxation,\cite{Nam2017La} as long as we are not too close to the magic angle. We assume that such triangular shape Fermi surface is relevant to the low-energy physics of the tBLG near the magic angle at $-1/2$ filling and base our analysis on this assumption. The key idea is that the almost parallel (well nested) sides of the triangular Fermi surfaces between $K$ and $K'$ valleys could lead to strong valley fluctuations, which further provides the  pairing glue for the superconductivity.

We describe a systematic procedure to extract an low-energy effective band structure from the continuum model described above. Briefly,  the end result is a single band model with the dispersion $\epsilon_{K,\vect{k}}=\vect{k}^2-\mu + \alpha (k_x^3-3k_x k_y^2)$ around the $K$ valley and $\epsilon_{K',\vect{k}}=\epsilon_{K,-\vect{k}}$ around the $K'$ valley.  In more detail, we proceed as follows. To model the triangular Fermi surface around the $\Gamma_M$ point, we first derive the effective band theory near $\Gamma_M$. One systematic and unbiased approach is to first collect the single-particle wave vectors $\ket{m\vect{k}}$ around $\Gamma_M$ in the middle band (including both its upper and lower branches), and then construct a density matrix $\rho\propto\sum_{m\vect{k}}\ket{m\vect{k}}\bra{m\vect{k}}$ out of these states (note that $\ket{m\vect{k}}$ are not orthogonal in the orbital space). By diagonalization $\rho=\sum_i\ket{\psi_i}p_i\bra{\psi_i}$, we can identify the leading natural orbitals $\ket{\psi_i}$ (orbitals with largest weights $p_i$). The number $n$ of the leading orbitals to be involved in the effective theory can be set by the desired fidelity level. To retain above 95\% fidelity, s.t.~$\sum_{i=1}^n p_i>0.95$, we typically need to take up to six orbitals (i.e.~$n=6$). Projecting the continuum model \eqnref{eq:Hcon} to the six orbitals leads to the effective Hamiltonian $H_K=\sum_{\vect{k}}c_\vect{k}^\dagger h_{K\vect{k}} c_\vect{k}$ with
\begin{equation}\label{eq:h6}
h_{K\vect{k}}=\mat{\epsilon_1 \sigma^1 & \kappa^{-}_{\vect{k}} & \kappa^{-}_{\vect{k}}\\
\kappa^{+}_{\vect{k}} & \lambda_{\vect{k}}& 0\\
\kappa^{+}_{\vect{k}} & 0 & -\lambda_{\vect{k}}}
\end{equation}
where $\kappa^{\pm}_{\vect{k}}=v_1(k_x \sigma^0 \pm \ii k_y \sigma^3)$ and $\lambda_{\vect{k}}=\epsilon_2+v_2 \vect{k}\cdot\vect{\sigma}$ are set by four real parameters $\epsilon_{1,2}$ and $v_{1,2}$. The band structure of the six-orbital model is shown in \figref{fig:bands}(b). We can see that the features around $\Gamma_M$ is well captured compared to the continuum model in \figref{fig:bands}(a), but the Dirac dispersions around $K_M$ and $K'_M$ can not be described by the six-orbital model (as expected). The six-orbital model provides a simpler and more flexible description of the near-$\Gamma_M$ band structure compared to the continuum model.\cite{Hejazi2018Mu} Its parameters can be determined by fitting to the first-principle calculations or experimental observations towards a more realistic modeling.

One can further simplify the six-orbital model by integrating out the high-energy electrons in the top and bottom bands, reducing the $6\times6$ Hamiltonian $h_{K\vect{k}}$ in \eqnref{eq:h6} to its first $2\times 2$ block: $h'_{K\vect{k}}=(\epsilon_1-b\vect{k}^2)\sigma^1+a\Re k_+^3 \sigma^0+\scO[k^4]$, which describes both branches of the middle band, where $k_\pm\equiv k_x\pm\ii k_y$ and the coefficients are given by $b=2\epsilon_1v_1^2/(\epsilon_2^2-\epsilon_1^2)$ and $a=4\epsilon_1\epsilon_2v_1^2v_2/(\epsilon_2^2-\epsilon_1^2)^2$. If we only focus on the lower branch, the effective band theory boils down to a single-orbital model
\begin{equation}\label{eq:H0}
\begin{split}
H_0&=\sum_{\vect{k}}c_{K\vect{k}}^\dagger \epsilon_{\vect{k}}c_{K\vect{k}}+c_{K'\vect{k}}^\dagger \epsilon_{-\vect{k}}c_{K'\vect{k}},\\
\epsilon_{\vect{k}}&=\vect{k}^2-\mu+\alpha \Re k_+^3,
\end{split}
\end{equation}
where we have chosen to rescaled the energy such that the single-orbital depends on only one tuning parameter $\alpha=a/b=2\epsilon_2v_2/(\epsilon_2^2-\epsilon_1^2)$ that characterizes the strength of the triangular Fermi surface anisotropy. The band structure of $\epsilon_{\vect{k}}$ is plotted in \figref{fig:bands}(c). In \eqnref{eq:H0}, the $K'$ valley Hamiltonian is also included, which can be inferred from that of the $K$ valley by the time-reversal symmetry $\scT:c_{K\vect{k}}\to\scK c_{K',-\vect{k}}$. The Fermi surfaces in both valleys are drawn in \figref{fig:pockets}(b) with $\mu=1, \alpha=1/3$ for example. One can see that the model essentially captures the triangular shape of the Fermi surface. There are three nesting vectors between $K$ and $K'$ pockets, which are set by the chemical potential $\mu$: $\vect{Q}_1=(\sqrt{3\mu},0)$ and $\vect{Q}_2=R_{2\pi/3}\vect{Q}_1$, $\vect{Q}_3=R_{-2\pi/3}\vect{Q}_1$ are related to $\vect{Q}_1$ by $C_3$ rotations. Note that the electronic spin degrees of freedom can be included in \eqnref{eq:H0} implicitly.

In this single-orbital model, the notions of  filling fraction and nesting commensurability are lost, but by going back to the original continuum model, we can identify the commensurate wavevector that has a high degree of nesting,  which is found to be the  $M_M$ points, i.e.~$\vect{Q}_{1}\simeq \vect{q}_2-\vect{q}_1/2$. 
A commensurate perfect nesting will be achieved at the filling $-5/8$, which is hole-doped by $25\%$ from the half-filling. We will show later in Sec.\,\ref{sec:IVCW} that including a commensurate inter valley ordering with a period corresponding to the $M_M$ point of the MBZ, we can induce a full gap for relatively small order parameters, and obtain an insulating state when we are at the filling $-(1/2+1/8)$ in the microscopic model given by the continuum theory \eqnref{eq:Hcon}. 

\subsection{Interactions and $\SO(4)$ Symmetry}\label{sec:interaction}

We now introduce interactions into the single-orbital model in \eqnref{eq:H0}. As the electron $c=(c_{K\uparrow},c_{K\downarrow},c_{K'\uparrow},c_{K'\downarrow})$ in the MBZ carries both the spin ($\sigma=\uparrow,\downarrow$) and the valley ($v=K,K'$) degrees of freedom, one may expect an emergent $\U(4)$ symmetry at low energy that rotates all four components of the electron, as pointed out in Ref.\,\onlinecite{Xu2018,Po2018,Fu2018,Zhang2018Low}. However, the electron kinetic energy (the band structure) strongly breaks this $\U(4)$ symmetry. For example, the triangular Fermi surface anisotropy $\alpha$ in the band Hamiltonian \eqnref{eq:H0} explicitly breaks the  symmetry as the Fermi surface deformations are opposite between the two valleys as shown in \figref{fig:pockets}. The $\U(4)$ symmetry is broken down to $\U(1)_c\times\U(1)_v\times\SO(4)$, where $\U(1)_c$ is the charge $\U(1)$ symmetry generated by $n_c=c^\dagger \sigma^{00} c$, $\U(1)_v$ denotes the emergent valley $\U(1)$ symmetry generated by $n_v=c^\dagger \sigma^{30} c$ and $\SO(4)\sim\SU(2)_{K}\times\SU(2)_{K'}$ stands for the two independent $\SU(2)$ spin rotation symmetries in both valleys generated by $\vect{S}_v=c_v^\dagger \vect{\sigma} c_v$ (for $v=K,K'$ separately). The original $\SU(4)$ generators that are broken by the Fermi surface anisotropy $\alpha$ form a (complex) $\SO(4)$ vector, which corresponds to the inter-valley coherence (IVC) order $I^\mu=c_K^\dagger s^\mu c_{K'}$ as proposed in Ref.\,\onlinecite{Po2018}, where $s^\mu$ are defined to be $(s^0,s^1,s^2,s^3)\equiv(\sigma^0,-\ii\sigma^1,-\ii\sigma^2,-\ii\sigma^3)$ for $\sigma^\mu$ being the Pauli matrices. The pairing  channels can also be classified by the $\SO(4)$ symmetry. There are only two possibilities: the inter-valley pairing $\Delta^\mu=c_K^\intercal \ii\sigma^2 s^\mu c_{K'}$ that transforms as $\SO(4)$ (pseudo)vector, and the intra-valley pairing $\Delta_v=c_v^\intercal \ii\sigma^2 c_v$ ($v=K,K'$) that transforms as $\SO(4)$ (pseudo)scalar. These operators are summarized in \tabref{tab:symmetry}, which exhaust all fermion bilinear operators that can be written down on a local Wannier orbital. (See the Appendix \ref{sec:AppA} for a more detailed classification of fermion bilinear operators.)

\begin{table}[htbp]
\caption{Symmetry classification of fermion bilinear operators (labeled in the bottom row). Electrical charge is labeled by  $q_{c}$, thus $q_c=0$ corresponds to charge neutral (particle-hole) operators, while $q_c=2$ corresponds to Cooper pair (particle-particle) operators. The valley quantum number of the $\U(1)_{v}$ symmetry  is labeled by $q_{v}$, hence inter-valley coherence order is obtained on condensing $q_v=2$ operators.  Non-Abelian symmetry representations are labeled by the dimension (with a prime to denote the pseudo- representation).}
\begin{center}
\begin{tabular}{|c|c|c|c|c|c|c|c|c|}
\hline
\multirow{4}{*}{$\U(4)$} & \multicolumn{2}{c|}{$\U(1)_c$} & \multicolumn{4}{c|}{$q_c=0$} & \multicolumn{2}{c|}{$q_c=2$}\\
\cline{2-9}
& \multicolumn{2}{c|}{$\SU(4)$} & $1$ & \multicolumn{3}{c|}{$15$}  & \multicolumn{2}{c|}{$6\oplus6'$}\\
\cline{3-9}
& $\simeq$ & $\U(1)_v$ & \multicolumn{3}{c|}{$q_v=0$} & $q_v=2$ & $q_v=0$ & $q_v=2$ \\
\cline{3-9}
& $\SO(6)$ & $\SO(4)$ & $1$ & $1'$ & $6$ & $4\oplus 4'$ & $4\oplus 4'$ & $2(1\oplus 1')$\\
\hline
\multicolumn{3}{c|}{}& $n_c$ & $n_v$ & $\vect{S}_v$ & $I^\mu$ & $\Delta^\mu$ & $\Delta_v$ \\
\cline{4-9}
\end{tabular}
\end{center}
\label{tab:symmetry}
\end{table}

Therefore any $\U(1)_c\times\U(1)_v\times\SO(4)$ symmetric local interaction should be mediated by one of these fermion bilinear channels. Further taken into account the time-reversal symmetry $\scT$ (that interchanges valleys), it turns out that there are only two linearly independent and symmetric interactions (see the Appendix \ref{sec:AppA} for the derivation of independent local interactions), which can be written purely in terms of density-density interactions as 
\begin{equation}\label{eq:Hint}
H_\text{int}=\sum_{\vect{q}}U_0 n_{K-\vect{q}}n_{K'\vect{q}}+\frac{U_1}{2}(n_{K-\vect{q}}n_{K\vect{q}} + n_{K'-\vect{q}}n_{K'\vect{q}}),
\end{equation}
where $n_{v\vect{q}}\equiv\sum_{\vect{k},\sigma}c_{v\sigma\vect{k}+\vect{q}}^\dagger c_{v\sigma\vect{k}}$ is the density operator of each valley. Since the density-density interaction is generally repulsive, we expect both parameters $U_0$ and $U_1$ to be positive (typically $U_0\approx U_1>0$). At the special point of $U_0=U_1=U$, the $\U(4)$ symmetry is restored for the interaction $H_\text{int}$. However, even if $H_\text{int}$ is tuned to the $\U(4)$ symmetric point, when combined with the kinetic energy $H_0$ in \eqnref{eq:H0}, the symmetry of the full Hamiltonian $H=H_0+H_\text{int}$ is still reduced to $\U(1)_c\times\U(1)_v\times\SO(4)$. Later in Sec.\,\ref{sec:breaking}, we will further discuss the effect of adding small interaction terms to finally break the emergent $\SO(4)$ symmetry down to the microscopic $\SO(3)$ spin rotation symmetry.

In summary, by putting together \eqnref{eq:H0} and \eqnref{eq:Hint}, we propose an effective model $H=H_0+H_\text{int}$ for the tBLG with Fermi level resting in the lower branch of the nearly-flat band but not too close to the charge neutrality (such that the Fermi surface is still within the control of $\Gamma_M$ point expansion). More specifically, we assume that the Fermi level does not go beyond the van Hove singularity that separates Fermi pockets around the $K_M$ points near charge neutrality from those centered around $\Gamma_M$, see also \figref{fig:pockets}(a).  Our remaining goal is to analyze the model within a weak coupling approach.

\subsection{Random Phase Approximation}\label{sec:RPA}

We calculate the renormalized interactions within the random phase approximation (RPA)\cite{Kuroki2008Unconventional,Graser2009Near,Maier2011} to analyze the electron instabilities in all six fermion bilinear channels as enumerated in \tabref{tab:symmetry}. We will first restrict our analysis within the $s$-wave channels for simplicity. For each fermion bilinear operator $A_{\vect{q}}=\frac{1}{2}\sum_{\vect{k}}\chi_{-\vect{k}+\vect{q}}^\intercal A\chi_{\vect{k}}$ generally expressed in the Majorana basis $\chi_\vect{k}$, we evaluate its bare static (zero frequency) susceptibility $\chi_{0}(\vect{q})=\langle A_{\vect{q}}^\dagger A_{\vect{q}}\rangle_0$ on the ground state of the single-orbital model $H_0$. Then we rewrite the interaction $H_\text{int}=g_0\sum_{\vect{q}}A_{\vect{q}}^\dagger A_{\vect{q}}+\cdots$ in the same channel to extract the bare coupling $g_0$. The RPA corrected coupling is then given by $g_\text{RPA}(\vect{q})=g_0 (1+g_0\chi_0(\vect{q}))^{-1}$. Admittedly, the RPA approach may not capture the interwind fluctuations in different channels. More systematic and unbiased approaches such as the function renormalization group\cite{Tang2018Sp} could be implemented to improve the result in the future. 

\begin{figure}[htbp]
\begin{center}
\includegraphics[width=0.74\linewidth]{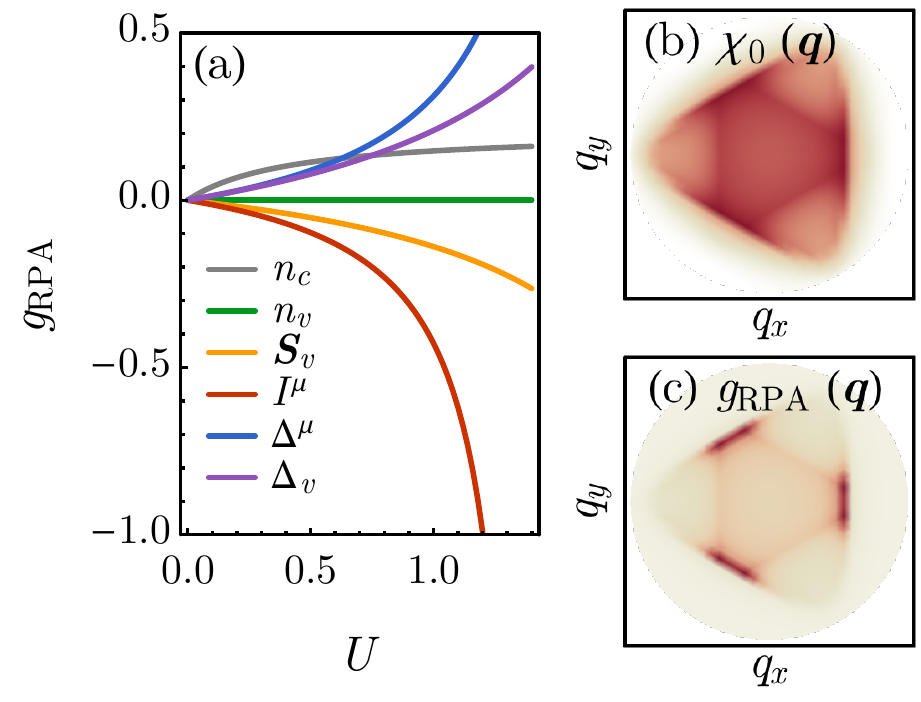}
\caption{(a) RPA effective coupling $g_\text{RPA}$ in different interaction channels v.s. the bare interaction strength $U_0=U_1=U$. The inter-valley coherence (IVC) channel $I^\mu$ has the strongest instability. (b) The bare susceptibility $\chi_0(\vect{q})=\langle I^{\mu\dagger}_\vect{q}I^\mu_\vect{q}\rangle_0$ of the IVC order at zero frequency ($\omega=0$). (c) The RPA corrected coupling $g_\text{RPA}(\vect{q})$ in the IVC channel. The coupling is strongly peaked around the nesting momentums.}
\label{fig:RPA}
\end{center}
\end{figure}

The largest (in magnitude) value of $g_\text{RPA}(\vect{q})$ is taken and plotted in \figref{fig:RPA}(a) as a function of $U_0=U_1=U$ for various channels. The most attractive coupling appears in the IVC channel, which is associated with the operator $I^\mu_\vect{q}=\sum_{\vect{k}}c_{K\vect{k}+\vect{q}}^\dagger s^\mu c_{K'\vect{k}}$. \figref{fig:RPA}(b) shows the bare susceptibility of the IVC fluctuation and \figref{fig:RPA}(c) is its RPA corrected coupling, which peaks strongly around three momentums that exactly correspond to the nesting momentums $\vect{Q}_{1,2,3}$. So as the bare interaction is strong enough, $I^\mu$ will condense at these momentums, leading to a finite-momentum IVC order, which we called the inter-valley coherence wave (IVCW). Suppose the nesting vector is pinned by the Moir\'e pattern to $M_M$.

Upon doping, the nesting condition will quickly deteriorate and the IVCW order will cease to develop. Nevertheless the low-energy valley fluctuations can play the role of the pairing glue, mediating an effective pairing interaction between electrons. A hint that can already be observed from \figref{fig:RPA} in which the attractive coupling diverges in the $I^\mu$ channel, while at the same time a repulsive coupling in the $s$-wave inter-valley pairing $\Delta^\mu$ channel also diverges. This implies that if the pairing form factor is allowed to change sign along the Fermi surface (which goes beyond $s$-wave), the repulsive coupling in this pairing channel can be effectively converted to an attractive one, leading to a strong pairing instability based on the Kohn-Luttinger mechanism\cite{Kohn1965Ne,Maiti2013Su}. The details will be discussed in the following.

\subsection{Superconductivity}\label{sec:TSC}

\begin{figure}[htbp]
\begin{center}
\includegraphics[width=0.98\linewidth]{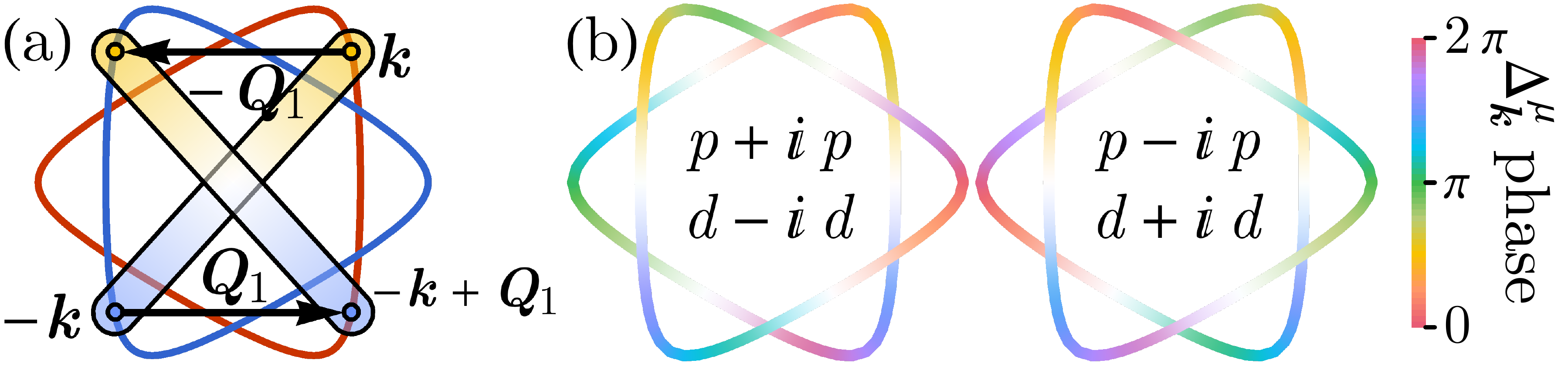}
\caption{(a) A Cooper pair scattered by the valley fluctuation of the nesting vector $\vect{Q}_1$ leads to a sign change along the Fermi surface (between $\Delta_{\vect{k}}^\mu$ and $\Delta_{-\vect{k}+\vect{Q}_1}^\mu$). (b) The leading inter-valley pairing form factors on the Fermi surface. The pairing phase is indicated by the hue and the gap size by the color intensity. Here we show the case of $w_d/w_p=1$ (i.e.~$d$-wave and $p$-wave are equal in strength) such that there are nodal points on the Fermi surface. For generic $w_d/w_p$, the Fermi surface will be fully gapped.}
\label{fig:pairing}
\end{center}
\end{figure}

To pin down the pairing instability mediated by the valley fluctuations, we take the RPA corrected interaction in the IVCW channel $I^{\mu\dagger} I^\mu$ and recast it in the inter-valley pairing channel $\Delta^{\mu\dagger}\Delta^\mu$ (restricting to the zero momentum pairing $c_{\vect{k}}c_{-\vect{k}}$)
\begin{equation}\label{eq:II=DD}
\sum_{\vect{q},\mu}g_\text{RPA}(\vect{q})I_{\vect{q}}^{\mu\dagger}I_{\vect{q}}^{\mu}\simeq -\sum_{\vect{q},\vect{k},\mu}g_\text{RPA}(\vect{q})\Delta^{\mu\dagger}_{-\vect{k}+\vect{q}}\Delta^{\mu}_{\vect{k}},
\end{equation}
where $I^\mu_\vect{q}=\sum_{\vect{k}}c_{K\vect{k}+\vect{q}}^\dagger s^\mu c_{K'\vect{k}}$ is the IVCW operator and $\Delta_{\vect{k}}^\mu=c_{K\vect{k}}^\intercal\ii\sigma^2 s^\mu c_{K'-\vect{k}}$ is the inter-valley pairing operator, recall that $(s^0,\vect{s})=(\sigma^0,-\ii\vect{\sigma})$. The attractive interaction ($g_\text{RPA}<0$) in the IVCW channel implies the repulsive interaction ($-g_\text{RPA}>0$) between $\Delta^{\mu}_{\vect{k}}$ and $\Delta^{\mu}_{-\vect{k}+\vect{q}}$. So the pairing can gain  energy only if there is a relative sign change between the pairing form factors connected by the nesting momentums $\vect{Q}_a$ (at which the scattering is the strongest), i.e.~$\Delta^{\mu}_{\vect{k}}=-\Delta^{\mu}_{-\vect{k}+\vect{Q}_a}$, as illustrated in \figref{fig:pairing}(a).

By solving the linearized gap equation, 
\begin{equation}
\sum_{\vect{k}'\in\text{FS}}v_F^{-1}(\vect{k}')g_\text{RPA}(\vect{k}+\vect{k}')\Delta_{\vect{k}'}^{\mu}=\lambda\Delta_{\vect{k}}^{\mu},
\end{equation}
the leading gap function (i.e.~the eigen function $\Delta_{\vect{k}}^\mu$ with the largest eigenvalue $\lambda$) is found to be of the form
\begin{equation}\label{eq:Delta}
\Delta_{\vect{k}}^\mu=u^\mu w_{\vect{k}} + v^\mu w_{\vect{k}}^*,\end{equation}
where $u^\mu$ and $v^\mu$ are complex vectors, and the form factor $w_{\vect{k}}=w_d k_+^2+w_p k_-$ is a linear combination of the $d+\ii d$ and the $p-\ii p$ waves with real coefficients $w_d$ and $w_p$, as shown in \figref{fig:pairing}(b). The mixing between the $d+\ii d$ and the $p-\ii p$ pairing is generic, because in the presence of the triangular Fermi surface distortion $\alpha$, the angular momentum is only mod 3 conserved, meaning that there is no distinction between the  $d+\ii d$ and the $p-\ii p$ wave on symmetry ground. The ratio $|w_p/w_d|$ carries the dimension of momentum and sets a momentum scale $k_Q=|w_p/w_d|$, which is expected to be associated with the nesting momentum $k_Q\simeq |\vect{Q}_a|/2$. The form factor $w_{\vect{k}}$ has three zeros (vortices) on the circle of $k_Q$ in the momentum space. If the Fermi surface circumvents the zeros from outside (or inside), the pairing will be dominated by $d+\ii d$ (or $p-\ii p$) wave.

{\bf Topological Superconductivity:} To determine the coefficients $u^\mu$ and $v^\mu$ in \eqnref{eq:Delta}, we can write down the Landau-Ginzburg (LG) free energy $F$ within the mean-field theory,\cite{Xu2018} (see also the Appendix \ref{sec:LG} for the derivation of Landau-Ginzburg free energy and a more detailed analysis of competing orders)
\begin{equation}
F=\sum_{\vect{k}}r \Delta_{\vect{k}}^{\mu*}\Delta_{\vect{k}}^{\mu}+\kappa (2(\Delta_{\vect{k}}^{\mu*}\Delta_{\vect{k}}^{\mu})^2-|\Delta_{\vect{k}}^{\mu}\Delta_{\vect{k}}^{\mu}|^2)+\cdots.
\end{equation}
As studied in Ref.\,\onlinecite{Xu2018}, the free energy admits two types of minimum, which are degenerated in energy,
\begin{equation}\label{eq:TSCsols}
\begin{split}
\text{chiral}&:
\left\{\begin{array}{rl}
u^\mu=&e^{\ii \phi} n^{\mu},\\
v^\mu=&0,
\end{array}\right.\text{ or }
\left\{\begin{array}{rl}
u^\mu=&0,\\
v^\mu=&e^{\ii \phi} n^{\mu},
\end{array}\right.\\
\text{helical}&:
\left\{\begin{array}{rl}
u^\mu=&e^{\ii \phi_1} (n_1^{\mu}+\ii n_2^{\mu}),\\
v^\mu=&e^{\ii \phi_2} (n_1^{\mu}-\ii n_2^{\mu}),
\end{array}\right.\\
\end{split}
\end{equation}
where $\phi,\phi_1,\phi_2$ are arbitrary phases and $n^\mu,n_1^\mu,n_2^\mu$ are real $\O(4)$ vectors with $n_1^\mu n_2^\mu=0$. The chiral solution preferentially choose the form factor of one chirality (either $w_{\vect{k}}$ or $w_{\vect{k}}^*$), which corresponds to four copies of the $d+\ii d$ or the $p-\ii p$ superconductors (or its time-reversal partners). The helical solution is a  superposition of $w_{\vect{k}}$ (in one spin sector) and $w_{\vect{k}}^*$ (in the other spin sector), which corresponds to two copies of the $d\pm\ii d$ or the $p\mp\ii p$ superconductors.

In the valley and spin space, $\Delta_{\vect{k}}^\mu$ transforms as a (complex) $\SO(4)$ vector, whose four components corresponds to the spin-singlet pairing $\Delta_{\vect{k}}^0$ and the spin-triplet pairing $\vect{\Delta}_{\vect{k}}=(\Delta_{\vect{k}}^1,\Delta_{\vect{k}}^2,\Delta_{\vect{k}}^3)$. In the presence of the emergent $\SO(4)$ symmetry, the singlet and triplet pairings are degenerated. This can be considered as an $\SO(4)$ generalization of the $\SO(3)$ pairing $\vect{\Delta}_{\vect{k}}$ proposed in Ref.\,\onlinecite{Xu2018}, such that the singlet pairing is also included as a possible option in our discussion. However, the $\SO(4)$ symmetry is not exact in the tBLG. Any inter-valley spin-spin interaction will break the $\SO(4)$ symmetry down to the global (valley-locked) $\SO(3)$ spin rotation symmetry, and thus splits the degeneracy between singlet and triplet pairings. If the singlet pairing is favored, then only the chiral gap function is possible, because there is no room for two perpendicular $\O(4)$ vectors $n_1^\mu$ and $n_2^\mu$ to coexist just in the singlet channel. If the triplet pairing is favored, then both the chiral and helical gap functions are allowed. We will discuss the effective of explicit $\SO(4)$ symmetry breaking in more details later.

In general, the superconductor will be a topological superconductor (TSC) with fully gapped Fermi surface.\cite{Qi2009,Fu2010,Qi2010} The chiral TSC breaks the time-reversal symmetry and also breaks the $\U(1)_c\times\U(1)_v\times\SO(4)$ symmetry to $\dsZ_2^F\times\U(1)_v\times\SO(3)$. The topological classification for the chiral TSC is $\dsZ$. If the $d+\ii d$ (or $p-\ii p$) pairing is stronger, the topological index will be $\nu=8$ (or  $\nu=-4$), which admits 8 (or 4) chiral Majorana edge modes. The helical (non-chiral) TSC preserves the (spin-flipping) time-reversal symmetry $\dsZ_2^\scT$ (under which $c_{K\vect{k}}\to \scK \ii\sigma^2 c_{K',-\vect{k}}, c_{K'\vect{k}}\to \scK \ii\sigma^2 c_{K,-\vect{k}}$) and breaks the $\U(1)_c\times\U(1)_v\times\SO(4)$ symmetry to $\dsZ_2^F\times\U(1)_v\times\SO(2)$. The $\SO(2)$ symmetry may be loosely called a spin $\U(1)_s$ symmetry since it corresponds to a joint spin rotation for both valleys (in either the same or the opposite manner). In the presence of both $\U(1)_v$ and $\U(1)_s$, the topological classification of the helical TSC is also $\dsZ$. If the $d\pm\ii d$ (or $p\mp\ii p$) pairing is stronger, the topological index will be $\nu=4$ (or  $\nu=-2$), which admits 4 (or 2) helical Majorana edge modes. It is also possible to fine tune the ratio $w_d/w_p$ to the topological phase transition between the $d$-wave and $p$-wave TSC, then superconducting gap will close at the nodal points on the Fermi surface resulting in 12 Majorana cones in the bulk.

{\bf Nematic Superconductivity:} Finally, we would like to briefly comment on the possibility of the nematic $d$-wave or $p$-wave pairing. We could go beyond the mean-field theory by considering more general momentum-dependent quartic terms in the LG free energy
\begin{equation}
\sum_{\vect{k},\vect{k}'}\kappa_{\vect{k}\vect{k}'}(2\Delta_{\vect{k}}^{\mu*}\Delta_{\vect{k}}^{\mu}\Delta_{\vect{k}'}^{\nu*}\Delta_{\vect{k}'}^{\nu}-\Delta_{\vect{k}}^{\mu*}\Delta_{\vect{k}}^{\mu*}\Delta_{\vect{k}'}^{\nu}\Delta_{\vect{k}'}^{\nu}).
\end{equation}
If $\kappa_{\vect{k}\vect{k}'}$ satisfies $\sum_{\vect{k},\vect{k}'}\kappa_{\vect{k}\vect{k}'}(w_{\vect{k}}^*w_{\vect{k}'})^2<0$, the LG free energy will have only one type of minimum, (see Appendix \ref{sec:LG} for justifications of the assumption and the solution)
\begin{equation}\label{eq:NSCsols}
\text{nematic:}\left\{\begin{array}{rl}
u^\mu=&e^{\ii \phi_1} n^{\mu},\\
v^\mu=&e^{\ii \phi_2} n^{\mu},
\end{array}\right.
\end{equation}
where $\phi_1$, $\phi_2$ are arbitrary phases and $n^\mu$ is a real $\O(4)$ vector. This solution corresponds to the nodal $d$-wave or $p$-wave pairing, as $\Delta_{\vect{k}}^\mu\sim\Re (e^{\ii(\phi_1-\phi_2)}w_{\vect{k}})n^\mu$, which preserves the time-reversal symmetry and breaks the $\U(1)_c\times\U(1)_v\times\SO(4)$ symmetry down to $\dsZ_2^F\times\U(1)_v\times\SO(3)$. The nodal line lies along the direction set by $\phi_1-\phi_2$, which breaks the $C_3$ rotational symmetry. So the nodal superconductor also has a ``nematic'' (orientational) order\cite{Grover2010Weak,Fu2014}. As the Fermi surface is not fully gapped, the nematic superconductor is not topological and has no protected edge mode. Apart from strong coupling, explicit breaking of $C_3$ rotation symmetry could also favor nematic superconductivity. 

\subsection{Slater Insulator and Valley Order}\label{sec:IVCW}

When the Fermi surface is tune to optimal nesting, the strong nesting instability could lead to the condensation of the IVC order parameter $I^\mu$ at the nesting momentums, which drives the system into the IVCW phase. In the weak coupling approach, the IVCW and the TSC order compete for the Fermi surface density of state. Here we provide a mean-field theory calculation that captures both competing orders and gives a rough estimate of the overall structure of the phase diagram. We start with the mean-field Hamiltonian $H_\text{MF}$ that incorporates the order parameters of both the IVCW $I_{\vect{Q}}^0$ and the TSC $\Delta_{\vect{k}}^0$ (which are restricted to the singlet channel without loss of generality given the $\SO(4)$ symmetry),
\begin{equation}\label{eq:HMF}
\begin{split}
H_\text{MF}&=H_0+g_I H_I+g_\Delta H_\Delta,\\
H_I&=\sum_{\vect{Q},\vect{k}}I_{\vect{Q}}^{0*}c_{K\vect{k}+\vect{Q}}^\dagger c_{K'\vect{k}}+h.c.+I_{\vect{Q}}^{0*}I_{\vect{Q}}^{0},\\
H_\Delta&=\sum_{\vect{Q},\vect{k}}\Delta_{\vect{k}}^{0*}c_{K\vect{k}}^\intercal\ii\sigma^2c_{K'-\vect{k}}+h.c.-\Delta_{-\vect{k}+\vect{Q}}^{0*}\Delta_{\vect{k}}^0,
\end{split}
\end{equation}
where $H_0$ is taken to be the single-orbital model \eqnref{eq:H0} and $\vect{Q}$ is summed over the three nesting vectors $\vect{Q}_{1,2,3}$. $g_I=g_\text{RPA}(\vect{Q})$ and $g_\Delta=\mathrm{avg}_{\vect{k},\vect{k}'\in\text{FS}}g_\text{RPA}(\vect{k}+\vect{k}')$ are the effective couplings in the IVC and the pairing channels respectively. Both of them originate from the RPA corrected coupling $g_\text{RPA}(\vect{q})$ and are expected to scale together with the interaction strength $U=U_0=U_1$. By tracing out the electron, we obtain the free energy $F=-\beta^{-1}\ln\Tr e^{-\beta H_\text{MF}}$ for the order parameters $I_{\vect{Q}}^0$ and $\Delta_{\vect{k}}^0$. (See the Appendix \ref{sec:SCMF} for more details on the self-consistent mean-field theory.) We find the free energy saddle point solution in the low temperature limit for different $W/U\sim g_I^{-1},g_\Delta^{-1}$ (where $W$ is the band width) and different chemical potentials $\mu$. This allows us to map out the mean-field phase diagram (in the zero temperature limit) as shown in \figref{fig:phases}. As we tune the twist angle towards the magic angle, the band gets flatten and the effective coupling increases. The TSC phase will first appear at low temperature. With stronger coupling, the IVCW phase will emerge around the optimal nesting and gradually expand to a wider range of chemical potential.

\begin{figure}[htbp]
\begin{center}
\includegraphics[width=0.9\linewidth]{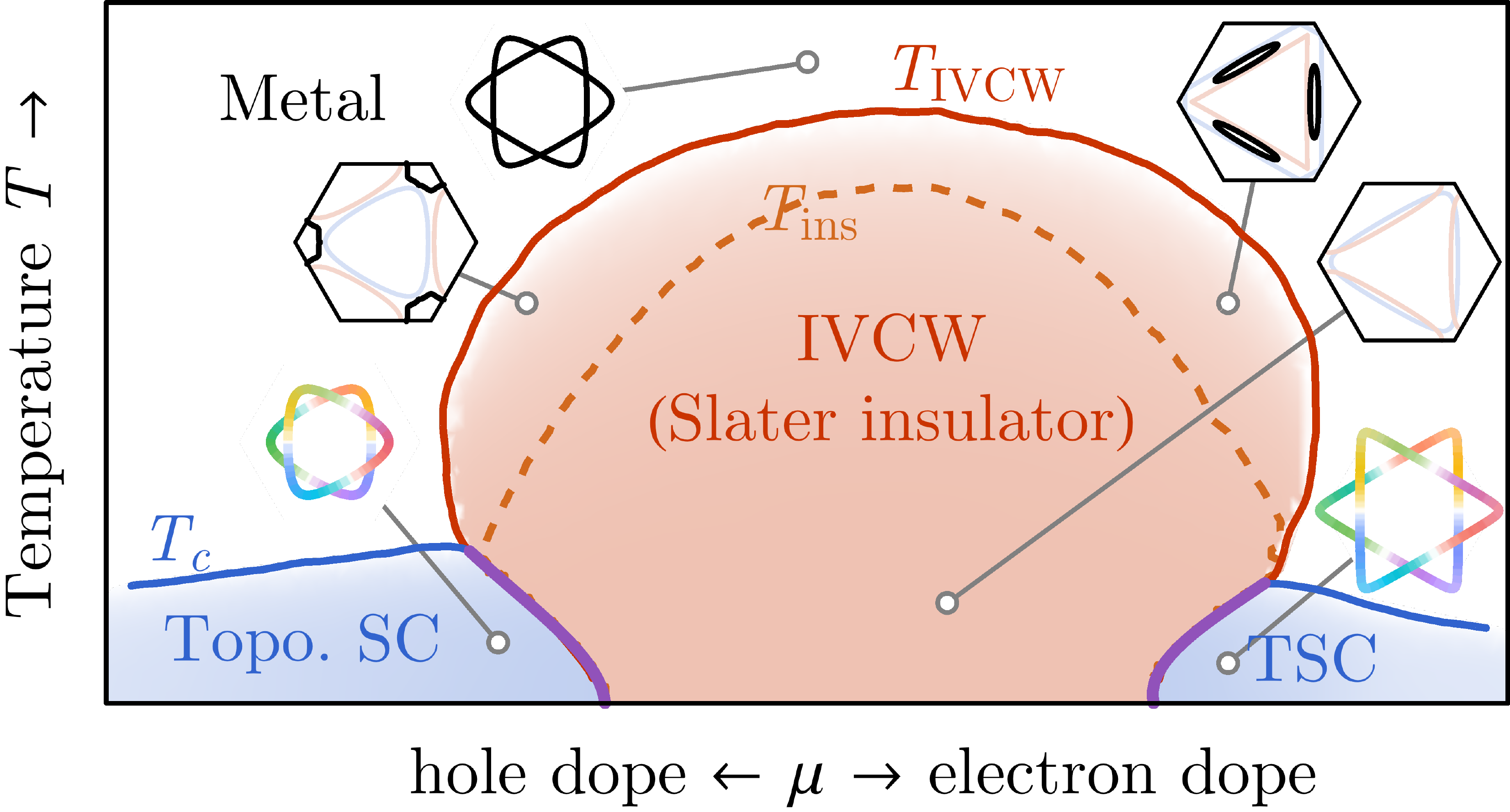}
\caption{Mean-field phase diagram in the vicinity of $f=-1/2$ and at finite temperature. TSC: topological superconductor, IVCW: inter-valley coherence wave. The TSC appears below $T_c$ around the IVCW insulator on both the hole and electron doped sides, with a $d+\ii d$ and $p-\ii p$ mixed inter-valley pairing. The IVCW order on sets at the temperature $T_\text{IVCW}$ and becomes strong enough to full gap out the Fermi surface below $T_\text{ins}$. On the hole doping side, the metallic IVCW phase has a single hole pocket with twofold spin degeneracy. The transition temperatures $T_c$ and $T_\text{IVCW}$ are correlated since they arise from the same interaction $g_\text{RPA}$.}
\label{fig:meanfield}
\end{center}
\end{figure}

As we fix the couplings at $g_I=0.8$ and $g_\Delta=0.4$ (the energy unit is set by the band dispersion in $H_0$), assume that the optimal nesting is around $\mu=1$ (such that the nesting momentum is $|\vect{Q}|=\sqrt{3\mu}\approx1.73$), and take the anisotropy parameter to be $\alpha=1/3$, we can obtain a mean-field phase diagram (for finite temperature) as shown in \figref{fig:meanfield} (by solving the free energy saddle point equations). The fermilogy at different representative points in the phase diagram are shown in \figref{fig:meanfield}. In the metallic phase, the Fermi surface consists of electron pockets around $K$ and $K'$ valleys (drawn together). In the TSC phase, the Fermi surface is gapped by the inter-valley pairing with the pairing form factor shown in color (following \figref{fig:pairing}(b)). The pairing can be either chiral or helical within the mean-field theory. In the IVCW phase, the $K'$ pocket (in light red) is shifted away from the $K$ pocket (in light blue) by the three nesting vectors $\vect{Q}_{1,2,3}$. Deep in the IVCW phase, the Fermi surface can be fully gapped. In between $T_\text{IVCW}$ and $T_\text{ins}$, small (reconstructed) hole or electron pockets remain on the Fermi level. However, using the single-orbital model \eqnref{eq:H0} as the starting point, we have lost track of the notion of the Moir\'e Brillouin zone (MBZ) and we can not tell if the nesting vector $\vect{Q}$ is commensurate with the Moir\'e lattice or not.

To further investigate the commensurability of the nesting vector and the corresponding filling of IVCW state, we have to fall back on the \emph{continuum model} \eqnref{eq:Hcon} for $H_0$, such that the MBZ can be referred. We would like to explore if the IVCW order can fully gap out the Fermi surface and lead to an insulator. We will first focus on the commensurate IVCW order. From the shape of the Fermi surfaces in \figref{fig:pockets}(a), the nesting vectors are most likely to be commensurate if they connect the $\Gamma_M$ point to the $M_M$ points in the MBZ. With this, we consider the IVCW order where the valley fluctuations simultaneously develops at the three $M_M$ points in the MBZ (corresponding to the nesting vector $\vect{Q}_1=\vect{q}_2-\vect{q}_1/2$ and its $C_3$ related partners $\vect{Q}_{2,3}$).

\begin{figure}[htbp]
\begin{center}
\includegraphics[width=0.75\linewidth]{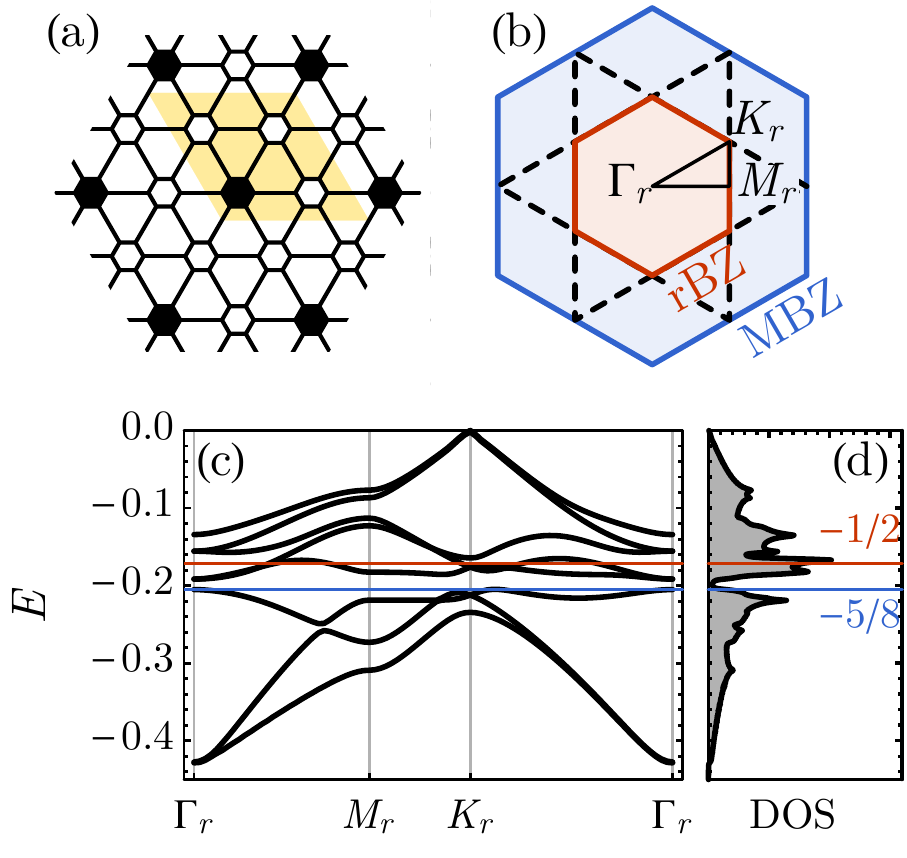}
\caption{(a) A $2\times2$ pattern on the Moir\'e lattice (little hexagons represent the AA stacking regions). The enlarged unit-cell is highlighted. (b) The reduced Brillouin zone (rBZ) compared to the Moir\'e Brillouin zone (MBZ). (c) The band structure of the IVCW state below neutrality. (d) The corresponding density of state (DOS) shows a full gap at filling $-5/8$.}
\label{fig:IVCW}
\end{center}
\end{figure}

The commensurate IVCW order breaks both the $\U(1)_v\times\SO(4)$ symmetry and the translation symmetry. It leads to a $2\times 2$ modulation on the Moir\'e lattice as demonstrated in \figref{fig:IVCW}(a). As the unit-cell is enlarges to four Moir\'e sites, the Brillouin zone will be reduced to 1/4 of the MBZ, as illustrated in \figref{fig:IVCW}(b). The lower branch of the band (from charge neutrality to the band bottom) will be folded to eight bands in the reduced Brillouin zone (rBZ), which consist of four folded bands for each valley. As we turn on the IVCW order to mix the $K$ and $K'$ valleys together, a full gap opens between the third and the fourth bands (counting from bottom) as shown in \figref{fig:IVCW}(c,d). Counting from the charge neutrality, this corresponds to the filling $f=-5/8$, but not the filling $f=-1/2$ as one may expect. In fact, the $-1/2$ level lies in the continuum above the IVCW gap, as indicated in \figref{fig:IVCW}(c,d). At the filling $-5/8$, the system becomes an IVCW ordered band insulator, which may be called a Slater insulator (to be distinguished from the Mott insulator). There is a simple geometric picture to explain the seemly strange $-5/8$ filling. In the ideal case, if we consider the $K$ and $K'$ pockets to be straight triangles connecting the $M_M$ points, illustrated as the dashed lines in \figref{fig:IVCW}(b), the nesting will be perfect at the desired $M_M$ momentum and the corresponding filling is indeed $-5/8$ by counting the areas of the triangles. Therefore, although the commensurate IVCW order can lead to a fully gapped insulator, but the filling of the insulator has a $1/8$ deficit from the $-1/2$ filling. We also checked that if the ordering momentum is changed to the $\Gamma_M$ or $K_M$ point momentum, no gap opening is observed with weak to medium IVCW order. While the $-5/8$ filling sounds peculiar, we note that in a recent experiment\cite{Yankowitz2018Tu} of tBLG, separate quantum oscillations (Landau fans) are observed to emerge from $f=-1/2$ and $f\approx -0.6$, the later of which is closer to $f=-5/8=-0.625$, although more evidences are still needed to verify or falsify this insulating state as a commensurate IVCW state.

However, if we go beyond the commensurate nesting and relax the nesting vector from the $M_M$ momentum, it is possible to obtain an incommensurate IVCW insulator for a range of fillings around $-5/8$, including the $-1/2$ filling typically, as long as the nesting condition is good. Another possibility is that the band structure may receive self-energy corrections from the interaction in such a way that the $-1/2$ filling Fermi surface turns out to admit good commensurate nesting. But in either picture, the $-1/2$ filling is not special compared to other fillings in terms of forming a Slater insulator, which still does not provide a natural explanation for the specific filling of the Mott insulator. This suggests that the Mott insulator in the tBLG might be a strongly correlated phase beyond the weak coupling picture like Fermi surface nesting. In this case, a strong coupling approach is required to understand the observed Mott insulator at precisely $-1/2$ filling. Below we discuss a scenario of Mott insulator that naturally arise from quantum disordering the adjacent superconducting phase by double-vortex condensation.\cite{Read1989, Kivelson1990, Sachdev1992,Balents1999,Senthil2000Z2}

\subsection{Mott Insulator and Topological Order}\label{sec:Mott}

One approach towards a strong-coupling Mott state is to start from the adjacent superconducting state and then suppress the $\U(1)_c$ charge fluctuation by proliferating double vortices of the superconductivity (SC) order parameter (or equivalently $2\pi$ fluxes seen by the electron).\cite{Read1989, Kivelson1990, Sachdev1992,Balents1999,Senthil2000Z2}
Single vortices of the SC order parameter become anyonic excitations in the resulting Mott state, such that the Mott phase acquires intrinsic topological order.\cite{Wen1990,Chen2010} In this approach, the nature of the topological order in the Mott phase will be closely related to the nature of the SC order in the adjacent SC phase. Here we assume that the nature of the SC order will remain qualitatively the same as we increase the interaction strength from the weak-coupling to the strong-coupling regime. This assumption is consistent with the past experience of unconventional superconductors including cuprates and iron-pnictides\cite{Dai2012Ma}. Assuming this, we can take the SC orders obtained from the weak-coupling approach as input to provide us with more insights about the possible orders in the Mott phase.

On the field theory level, this amounts to first fractionalizing the electron $c_{v\sigma}$ into a bosonic parton $b$ and a fermionic parton $f_{v\sigma}$ as $c_{v\sigma}=bf_{v\sigma}$ following a slave-boson approach\cite{Read1983,Coleman1984,Kotliar1986,Wen1991}, where $v=K,K'$ labels the valley and $\sigma=\uparrow,\downarrow$ labels the spin. Both bosonic and fermionic partons couple to the emergent gauge field. We assign the $\U(1)_c$ symmetry charge to the bosonic parton and the $\U(1)_v\times\SO(4)$ symmetry charge to the fermionic parton, in close analogy to the spin-charge separation in cuprates\cite{Rice1993,Anderson2004,Lee2006}. The fermionic parton is assumed to be in one of the SC state, such that once the bosonic parton condenses, the electronic SC state will be recovered. As we go from the (electronic) SC phase to the Mott phase, the bosonic parton is expected to acquire a gap across the transition, such that the charge fluctuations will be gapped and the $\U(1)_c$ symmetry will be restored in the Mott phase. Then the fermionic parton SC state essentially becomes a (generalized version of) quantum spin liquid with intrinsic topological order and symmetry fractionalization\cite{Yao2010,Levin2012Fractional,Essin2013,Barkeshli2014,Chen2015} of valley and spin quantum numbers. Hence such a Mott state may be called a valley-spin liquid (VSL). Different types of 
SC states correspond to different types of Mott states, as summarized in \tabref{tab:Mott}. On the other hand, charge doping the VSL states will drive the bosonic parton condensation $\langle b\rangle\neq0$, which identifies the fermionic parton $f_{v\sigma}$ with the electron $c_{v\sigma}=\langle b\rangle f_{v\sigma}$, and converts the topological order back to the corresponding SC order. So the correspondence between the SC states and the Mott states in \tabref{tab:Mott} are mutually consistent.

\begin{table}[htbp]
\caption{Possible Mott states originated from adjacent SC states.}
\begin{center}
\begin{tabular}{|c|c||c|c|}
\hline
\multicolumn{2}{|c||}{SC phase} & \multicolumn{2}{c|}{Mott phase} \\ 
\hline
type & pairing & state & symmetry\\
\hline
\multirow{2}{*}{chiral} & $d+\ii d$ & $\SO(8)_1$ VSL & $\U(1)_c\times\U(1)_v$\\
\cline{2-3}
& $p-\ii p$ & $\SO(4)_{-1}$ VSL & $\times\SO(3)$\\
\hline
\multirow{2}{*}{helical} & $d\pm\ii d$ & $\dsZ_2$ VSL + BSPT &  $\U(1)_c\times\U(1)_v$\\
\cline{2-3}
& $p\mp\ii p$ & $\dsZ_2$ VSL (SET) & $\times\U(1)_s\times\dsZ_2^{\scT}$\\
\hline
\multirow{2}{*}{nematic} & \multirow{2}{*}{$d$ or $p$} & gapless $\dsZ_2$  VSL & $\U(1)_c\times\U(1)_v$\\
&& + nematic order & $\times\SO(3)\times\dsZ_2^\scT$, \crossout{$C_3$}\\
\hline
\end{tabular}
\end{center}
\label{tab:Mott}
\end{table}

The chiral VSL sate can be viewed as the $d+\ii d$ (or $p-\ii p$) chiral TSC state of the fermionic parton, which enjoys the $\SO(8)_1$ (or $\SO(4)_{-1}$) topological order.\cite{Kitaev2003} They admit Abelian Chern-Simon theory\cite{Read1990,Blok1990,Wen1992,Lu2012,Hung2013Kmatrix} descriptions $\scL_\text{CS}=\frac{1}{4\pi} K_{IJ} a^I\wedge \dd a^J$ with the $K$ matrices given by
\begin{equation}
K_{\SO(4)_{-1}}=\mat{-2&0\\0&-2}, K_{\SO(8)_1}=\smat{2&-1&-1&-1\\-1&2&0&0\\-1&0&2&0\\-1&0&0&2}.
\end{equation}
Both topological orders have four anyon sectors, labeled by $\sfid$, $\sfe$, $\sfm$ and $\sfeps$. In the $\SO(4)_{-1}$ topological order state, $\sfe$ and $\sfm$ anyons are semions: one carries spin-1/2 (the projective representation of $
\SO(3)$) and no valley charge (the $\U(1)_v$ charge), the other carries valley  charge $\pm1$ and spin-0. They fuse to the fermionic spinon $\sfeps$ that carries both spin-1/2 and valley charge. This symmetry fractionalization pattern can be infer from the fact that the $\pi$-flux core in the $p-\ii p$ TSC traps 4 Majorana zero modes $\chi_{1,2,3,4}$, which splits into two sectors (differed by fermion parity) under the four-fermion interaction $H=V\chi_1\chi_2\chi_3\chi_4$, and the $\U(1)_v$ and $\SO(3)$ acts separately in either one of the sectors.\cite{You2015} After gauging the fermion parity, the two sectors are promoted to $\sfe$ and $\sfm$ anyons respectively. In the $\SO(8)_1$ topological order state, $\sfe$, $\sfm$, $\sfeps$ are all fermions. $\sfm$ carries no symmetry charge (because now the $\pi$-flux core traps 8 Majorana zero modes, which can be trivialized by the interaction in the even fermion parity sector), but $\sfe$ carries the same symmetry charges as the fermionic spinon $\sfeps$. The chiral VSL states are characterized by their non-trivial chiral central charges: $c=-2$ for $\SO(4)_{-1}$ and $c=4$ for $\SO(8)_1$. In the ideal case, the chiral central charge can be detected from the thermal Hall conductance as $\kappa_\text{H}=c \pi k_B^2 T/(6\hbar)$.\cite{Kane1997,Read2000,Gromov2015}

Now we turn to the helical VSL states, corresponding to the helical TSC states of fermionic partons. Both the $d$-wave and the $p$-wave parton TSC states lead to the $\dsZ_2$ topological order (described by the $K$ matrix $K_{\dsZ_2}=\smat{0&2\\2&0}$).\cite{Kou2008} Their difference lies in a topological response of the $\U(1)_v\times\U(1)_s$ symmetry, which might be called the valley-spin Hall conductance $\sigma_\text{vsH}$, defined as the coefficient in the following the effective response theory\cite{Qi2008,Ye2013,Cheng2014}
\begin{equation}
\scL[A_v,A_s]=\frac{\sigma_\text{vsH}}{2\pi}A_v\wedge\dd A_s,
\end{equation}
where $A_v$ and $A_s$ are the background fields that probe the $\U(1)_v\times\U(1)_s$ symmetry. The $\dsZ_2$ topological order have four anyon sectors: $\sfid$, $\sfe$, $\sfm$ and $\sfeps$, where $\sfe$ and $\sfm$ are bosons with mutual-semionic statistics, and they fuse to the fermionic parton $\sfeps$. For the $p$-wave helical VSL, $\sfe$ and $\sfm$ must separately carry either the $\U(1)_v$ or the $\U(1)_s$ symmetry charge, and $\sfeps$ carries both charges. The mutual-semionic statistics between $\sfe$ and $\sfm$ implies that the $p$-wave helical VSL state will have a fractionalized valley-spin Hall conductance $\sigma_\text{vsH}=-1/2$. Moreover, because the fermionic spinon $\sfeps$ is a Kramers doublet ($\scT^2=-1$) under the time-reversal symmetry,\footnote{In the presence of $\U(1)_v$, it is always possible to (re)define a time-reversal operation that acts on the fermionic parton as $\scT^2=-1$.} it must be the case that one of $\sfe$ or $\sfm$ is a Kramers doublet and the other one is a Kramers singlet ($\scT^2=+1$), such that the time-reversal anomaly vanishes\cite{Barkeshli2016,Wang2017}. So the $p$-wave helical VSL state is a $\U(1)_v\times\U(1)_s\times\dsZ_2^\scT$ symmetry\footnote{We will use the direct product notation loosely, because $\dsZ_2^\scT$ here can have several choices, whose relation to the $\U(1)_v$ and $\U(1)_s$ groups can vary. A canonical choice will be $(\U(1)_v\rtimes\dsZ_2^\scT)\times\U(1)_s$} enriched topological (SET) state\cite{Hung2013,Mesaros2013,Lu2016}.  For the $d$-wave helical VSL, $\sfm$ can be charge neutral and Kramers singlet, whereas $\sfe$ and $\sfeps$ both carry the $\U(1)_v\times\U(1)_s$ charge and are Kramers doublet. This can be viewed as a trivial $\dsZ_2$ topological order on top of a $\U(1)_v\times\U(1)_s$ bosonic symmetry protected topological (BSPT) state.\cite{Chen2011,Chen2012,Levin2012,Lu2012,Chen2013,You2014,Bi2017} The $\dsZ_2$ topological order can be removed by condensing the charge neutral boson $\sfm$. Then the Mott insulator simply realizes a $\U(1)_v\times\U(1)_s$ BSPT state with quantized valley-spin Hall conductance $\sigma_\text{vsH}=1$.

Finally, if we start with the nematic superconductor, the corresponding Mott state will be a gapless $\dsZ_2$ VSL with nodal fermionic partons and gapped visons.\cite{Grover2010Weak} The symmetry of this VSL state is $\U(1)_c\times\U(1)_v\times\SO(3)\times\dsZ_2^\scT$. Like the nematic superconductor, the $C_3$ rotation symmetry is still broken in the VSL state, so there will be a coexisting nematic order in this Mott insulator.

In all cases, the emergent $\SO(4)$ symmetry is broken in the Mott phase. But the remaining symmetry is still sufficient to protect a two-fold degeneracy of the electron. For the chiral VSL, the electron transforms (projectively) as spin-1/2 (spinor representation) of the $\SO(3)$ symmetry. For the helical VSL, the electron forms Kramers doublet under the time-reversal symmetry. For the nematic VSL, both $\SO(3)$ and time-reversal protections are present. The symmetry protected two-fold degeneracy in the valley-spin space is consistent with the experimentally observed Landau fan\cite{Cao2018a} near the Mott phase with the filling-factor sequence $2,4,6,\cdots$. Consider for example, the spin singlet  VSL phase, which is connected to the spin singlet chiral superconductor. Here, spin degeneracy is present, and although valley remains a good quantum number, since the phase itself breaks time reversal symmetry, the degeneracy between opposite valleys is lost. Although it is hard to estimate the strength of this effect, the symmetry dictated degeneracy is just twofold.

\subsection{Breaking SO(4) Symmetry}\label{sec:breaking}

Both the IVCW and the TSC phases break the emergent $\SO(4)$ symmetry, as their order parameters $I^\mu$ and $\Delta^\mu$ are  $\SO(4)$ vectors. The four (complex) components of the order parameters correspond to the orderings in the spin-singlet and the spin-triplet channels, which are degenerated in the presence of the $\SO(4)$ symmetry. However, the $\SO(4)$ symmetry is never exact in reality. The explicit $\SO(4)$ symmetry breaking can split the degeneracy. We will analyze the effects of the $\SO(4)$ symmetry breaking in the following.

We first consider the Heisenberg spin-spin interaction between valleys,
\begin{equation}
H_J= \sum_{\vect{q}} J(\vect{q})\vect{S}_{K\vect{q}}\cdot\vect{S}_{K'-\vect{q}},
\end{equation}
where $\vect{S}_{v\vect{q}}=\sum_{\vect{k}}c_{v\vect{k}+\vect{q}}^\dagger \vect{\sigma} c_{v\vect{k}}$ (for $v=K,K'$) is the spin operator. The $J(\vect{q})<0$ (or $J(\vect{q})>0$) case corresponds to the Hunds (or anti-Hunds) interaction. It belongs to the $(1,1)$ representation (the symmetric rank-2 tensor) of the $\SO(4)\simeq\SU(2)_K\times\SU(2)_{K'}$ group, which locks the two $\SU(2)$ subgroups together and breaks the $\SO(4)$ symmetry down to $\SO(3)$. The interaction $H_J$ admits decompositions in the IVC and the pairing channel as
\begin{equation}\label{eq:HJ}
\begin{split}
H_J &\simeq \frac{1}{8}\sum_{\vect{k},\vect{q}}J(\vect{q})(-3\Delta_{\vect{k}+\vect{q}}^{0\dagger}\Delta_{\vect{k}}^{0}+\vect{\Delta}_{\vect{k}+\vect{q}}^{\dagger}\cdot\vect{\Delta}_{\vect{k}})+\cdots,\\
&\simeq \frac{1}{8}\sum_{\vect{q}',\vect{q}}J(\vect{q}')(-3I_{\vect{q}}^{0\dagger}I_{\vect{q}}^{0}+\vect{I}_{\vect{q}}^{\dagger}\cdot\vect{I}_{\vect{q}})+\cdots,
\end{split}
\end{equation}
where $\Delta^0$ and $I^0$ are the spin-singlet orderings (as $\SO(3)$ scalar), and $\vect{\Delta}$ and $\vect{I}$ are the spin-triplet orderings (as $\SO(3)$ vector). Depending on the sign of the inter-valley Heisenberg interaction $J(\vect{q})$, the spin-triplet (or spin-singlet) pairing is favored if the interaction is Hunds (or anti-Hunds) like. If we assume an anti-Hunds interaction (i.e.~$J(\vect{q})>0$), then according to \eqnref{eq:HJ}, the interactiont will provide attractive interactions for both the IVC and the pairing in the spin-singlet channel. The anti-Hunds interaction could arise from the renormalized Hubbard interaction by integrating out high energy electrons as proposed in Ref.\,\cite{Chakravarty1991,Kivelson2018}. Note that the spin-singlet TSC can only be a chiral TSC as discussed in Sec.\,\ref{sec:TSC} previously. However, if the inter-valley interaction turns out to be Hunds like, the spin-triplet pairing could also be favored, which admits both chiral and helical TSC. The possibilities are summarized in \tabref{tab:orders}.


\begin{table}[htbp]
\caption{Orders favored by different interactions (marked by $\checkmark$). IVCW: inter-valley coherence wave, TSC: (inter-valley) topological superconductivity ($d+\ii d/p-\ii p$-wave), $s$-SC: (inter-valley) $s$-wave superconductivity. $I^0$ and $\Delta^0$ are in the spin-singlet channel, $\vect{I}$ and $\vect{\Delta}$ are in the spin-triplet channel.}
\begin{center}
\begin{tabular}{|c|c|c|c|c|c|c|c|}
\cline{3-8}
\multicolumn{2}{c|}{} & \multicolumn{2}{c|}{IVCW} & \multicolumn{2}{c|}{TSC} & \multicolumn{2}{c|}{$s$-SC}\\ \hline
\multicolumn{2}{|c|}{interaction} & $I^0$ & $\vect{I}$ & $\Delta^0$ & $\vect{\Delta}$ & $\Delta^0$ & $\vect{\Delta}$\\ \hline
\multicolumn{2}{|c|}{$\SO(4)$ symmetric} & $\checkmark$  & $\checkmark$ & $\checkmark$ & $\checkmark$ & & \\ \hline
\multicolumn{2}{|c|}{$+\vect{S}_{K}\cdot\vect{S}_{K'}$} & $\checkmark$ & & $\checkmark$ & & & \\ \hline
\multicolumn{2}{|c|}{$-\vect{S}_{K}\cdot\vect{S}_{K'}$} & & $\checkmark$ & & $\checkmark$ & & \\ \hline
\multirow{2}{*}{$-I^{0\dagger}I^{0}$}& weak & $\checkmark$ & & & $\checkmark$ & & \\ \cline{2-8}
& strong & $\checkmark$ & & & & $\checkmark$ & \\ \hline
\end{tabular}
\end{center}
\label{tab:orders}
\end{table}

However, if the $\SO(4)$ symmetry breaking is implemented in the IVC channel, the result can be very different. Suppose we consider the following enhanced attraction (i.e.~$g(\vect{q})<0$) in the $I^{0}$ channel, so as to single out the spin-singlet IVCW order. The same interaction would be translated into the pairing channel as
\begin{equation}\label{eq:Hg}
\begin{split}
&H_g=\sum_{\vect{q}}g(\vect{q})I_{\vect{q}}^{0\dagger}I_{\vect{q}}^{0}\\
&\simeq\frac{1}{2}\sum_{\vect{k},\vect{q}}g(\vect{q})(\Delta_{-\vect{k}+\vect{q}}^{0\dagger}\Delta_{\vect{k}}^{0}-\vect{\Delta}_{-\vect{k}+\vect{q}}^{\dagger}\cdot\vect{\Delta}_{\vect{k}})+\cdots,
\end{split}
\end{equation}
which is also an attractive interaction in the spin-singlet pairing channel $\Delta^0$ (note that $g<0$). In contrast to \eqnref{eq:II=DD}, only the $I^{0\dagger}I^0$ interaction is involved in \eqnref{eq:Hg}, which completely changes the interaction sign in the singlet pairing channel. Under the RPA correction, $g(\vect{q})$ peaks strongly around the nesting momentums $\vect{q}=\vect{Q}_{1,2,3}$, thus the attractive interaction between $\Delta_{\vect{k}}^0$ and $\Delta_{-\vect{k}+\vect{Q}_a}^0$ effectively reduces the energy gain in the singlet channel, due to the sign-changing TSC pairing form factor (i.e.~$\Delta_{\vect{k}}^0=-\Delta_{-\vect{k}+\vect{Q}_a}^0$). Therefore a slightly enhanced attractive interaction in the spin-singlet IVCW channel will actually suppresses the spin-singlet TSC pairing and favors the spin-triplet TSC pairing, as summarized in \tabref{tab:orders}. The spin-triplet TSC can be either chiral or helical as discussed Sec.\,\ref{sec:TSC} previously. Although $H_J$ in \eqnref{eq:HJ} and $H_g$ in \eqnref{eq:Hg} are both $\SO(4)$ symmetry breaking terms in the $(1,1)$ representation that favor the singlet IVCW order, yet their effects on splitting the singlet-triplet degeneracy in the TSC channel is completely opposite. This has to do with the fact that under the RPA correction, the interaction $H_J$ in the spin channel is not sensitive to the nesting effect, but the interaction $H_g$ in the valley channel exhibit a strong nesting effect. This results in  very different momentum-dependence of their coupling functions ($J(\vect{q})$ or $g(\vect{q})$), which finally divide the fate of the singlet-triplet splitting. The competition between these two symmetry breaking effects demands further analysis by more refined approach such as the functional renormalization group\cite{Honerkamp2001,Wang2009}, which will be left for future works. \cite{Kennes2018}

Finally, we would like to comment on the connection to Ref.\,\onlinecite{Kivelson2018}, where the valley XY interaction $H_g$ in \eqnref{eq:Hg} was considered to be the dominant interaction in the tBLG. In this case, the emergent $\SO(4)$ symmetry is strongly broken. The effective attraction in the spin-singlet pairing channel can simply drive the $s$-wave valley-symmetric spin-singlet pairing, which then leads to a nontopological superconductor as in ,\onlinecite{Kivelson2018}. Therefore, whether the superconductivity in the tBLG is topological or not could sensitively depend on the form and the strength of the $\SO(4)$ symmetry breaking interactions, as summarized in \tabref{tab:orders}.

\subsection{Effect of Electric Field}

\begin{figure}[htbp]
\begin{center}
\includegraphics[width=0.8\linewidth]{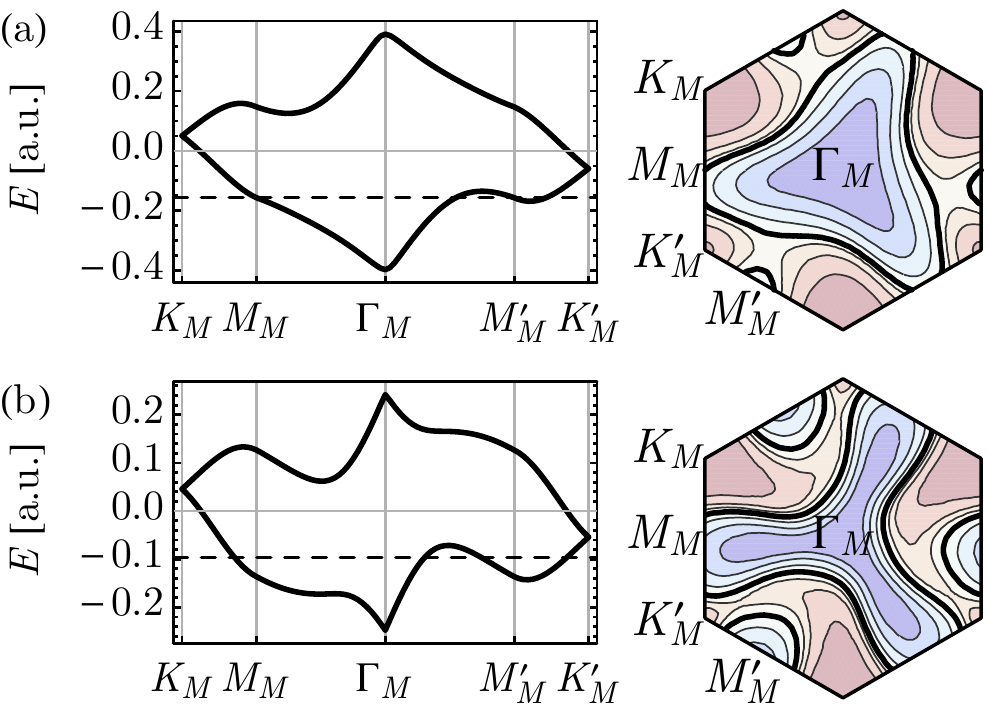}
\caption{Band structure (left panel) and the equal-filling Fermi surfaces (right panel) in the Moir\'e Brillouin zone around the $K$ valley in the presence of vertical electric field, for (a) weak field and (b) intermediate field. The $f=-1/2$ filling level is marked out as dashed lines in the band structure plot and as thick lines in the Fermi surface contour.}
\label{fig:Efield}
\end{center}
\end{figure}

Within the framework of the weak coupling theory, we can further consider the effective of a vertical electric field. In the continuum model, turning on the electric field amounts to introducing a potential difference between the layers,
\begin{equation}
H_K\to H_K+V\sum_{\vect{k}l} (-)^l c_{K_{l}\vect{k}}^\dagger c_{K_{l}\vect{k}}.
\end{equation}
As the time-reversal symmetry $\scT$ remains unbroken under the electric field, the $K'$ valley Hamiltonian $H_{K'}=\scT H_{K}\scT^{-1}$ is still related to that of the $K$ valley $H_{K}$ by the time-reversal operation. We will focus on the band structure around the $K$ valley. \figref{fig:Efield} shows the effect of the electric field on the band structure and the Fermi surfaces for the cases of (a) $V=0.2 v_F|\vect{q}_a|$ and (b) $V=0.6 v_F|\vect{q}_a|$. One can see that the Fermi surface is distorted as the electric field shifts the Dirac cones at $K_M$ and $K'_M$ relative to each other in energy (as they originated from the top and the bottom layers respectively).

We can follow the procedure described in Sec.\,\ref{sec:band} to extract the effect of the vertical electric field in the single-orbital model. However a symmetry analysis already suffices to determine the more relevant deformation of the Fermi surface. Given that the electric field breaks the $M_y:k_+\to k_-$ mirror symmetry and preserves the $C_3:k_+\to e^{2\pi\ii/3}k_+$ rotational symmetry, new terms can be added to the single-orbital model \eqnref{eq:H0} as
\begin{equation}\label{eq:Edeform}
\epsilon_{\vect{k}}\to\epsilon_{\vect{k}}+\alpha' \Im k_+^3+\alpha''\Im k_+^6+\cdots.
\end{equation}
The $\alpha'$ and $\alpha''$ terms describes the rotation and deformation of the Fermi surface as shown in \figref{fig:Efield}(a) for weak field. If the electric field is of the same order of the band width, the Fermi surface could be strongly deformed as in \figref{fig:Efield}(b), which goes beyond the perturbative description of \eqnref{eq:Edeform}.

As a consequence of the Fermi surface deformation, the Fermi surface nesting between $K$ and $K'$ valley will be suppressed by the electric field, therefore both the IVCW and the SC instability should reduce with the electric field. However as the deformation effect $\alpha''$ enters at a higher order perturbation, one expects that nesting-driven orders remains insensitive to weak electric field, until the interlayer electric potential difference $V$ reaches the order of the band width. Additional effects of interlayer electric field such as modulation of substrate effects due to the vertical displacement of the 2D electron gas can also play a role, and were not included in this analysis.  

\section{Discussion}\label{sec:summary}

In summary, we presented a weak coupling analysis of  valley fluctuation mediated superconductivity in  twisted bilayer graphene. We started with a momentum space formalism of the low-energy effective Moir\'e band structure, so as to circumvent the obstruction to constructing valley symmetric Wannier tight binding models. We identified the triangular (three-fold) anisotropy of the Fermi surface is a universal feature of the Moir\'e band structure around the charge neutrality, as it is the lowest-order distortion that is consistent with all the lattice symmetries. The Fermi surface anisotropy has important implications. The triangular shape of the Fermi surface allows a unique nesting between the parallel triangle sides of opposite valley Fermi pockets. This leads to enhanced valley fluctuations near half-filling, which in turn can provide the pairing glue which is demonstrated using the RPA.

By solving the pairing gap equation with the RPA corrected interaction, we obtain the leading pairing instability in the inter-valley channel with a $d+\ii d$ and $p-\ii p$ mixed pairing form factor. The mixing between the $d$-wave and $p$-wave pairing is generic, because with triangular anisotropy, and the remaining $C_3$ symmetry, the angular momentum of the electron is only  conserved modulo three, so there is no distinction between $d+\ii d$ and $p-\ii p$ on symmetry ground. Further taking spin into account, one obtains both spin singlet and triplet chiral superconductors, parameterized by a four-vector $n^\mu$, where the $\mu=0$ component corresponds to the spin-singlet. Additionally, helical pairing orders were also discussed, parameterized by two orthogonal $n^\mu$ vectors. 


We emphasized the approximate SO(4) spin-valley symmetry. The naive $\SU(4)$ symmetry of four component electrons in valley-spin space is broken by the Fermi surface distortion which is opposite between the two valleys, leading to $\U(1)_v\times\SO(4)$ symmetry. The $\SO(4)$ symmetry allows us to discuss the spin-singlet and spin-triplet pairings on equal footing, and $n^\mu$ transforms as an $\SO(4)$ vector. This degeneracy is lifted by $\SO(4)$ breaking perturbations and we argued that a Hunds (anti-Hunds) interaction, i.e. an inter-valley ferromagnetic (antiferromagnetic) spin interaction favors spin-triplet (spin-singlet) pairing, which can be probed by studying the response to a Zeeman field.  The presence of an approximate SO(4) symmetry could still have observable consequences which would be interesting to explore further. For example, if the SO(4) breaking is not too strong, a Zeeman field would tune a transition between singlet and triplet superconductors at low temperatures.  Thus twisted bilayer graphene may provide an opportunity to study different SC phases and the phase transitions between them. 


We propose two scenarios for the insulating phases. First,  pushing to stronger interactions we see that  inter valley coherence order can develop at the nesting vectors. The close commensurate wavevectors are the three $M$ points  corresponding to $(0,\,\pi)$, $(\pi,\,0)$ and $(\pi,\,\pi)$  at the midpoint of the  triangular lattice Brillouin Zone edges. The simultaneous condensation of  IVC order at these three wavevectors leads to a Slater insulator, although in our model a full gap obtains slightly below half filling at $f=-\frac 12 -\frac18$. Future work should establish if a more complete treatment of interactions changes this conclusion. Nevertheless, other aspects of the phenomenology appear promising. For example, on hole doping the IVCW insulator, a single Fermi pocket appears, with two fold degeneracy (see \figref{fig:meanfield}). This agrees with the observed quantum oscillation experiments on the hole doped side of the Mott insulator below neutrality, where a Landau fan degeneracy in multiples of two was observed. As with superconductivity, the SO(4) symmetry implies a degeneracy between spin singlet and spin triplet IVCW orders, the latter being a kind of spin density wave. The same Hunds (``anti-Hunds'' ) SO(4)  breaking interaction also picks out the spin-triplet (spin-singlet) IVCW order.

In Ref.\,\onlinecite{Cao2018}, superconductivity was found to coexist with the insulating phase, i.e.~superconducting puddles form even at half filling and establish a phase coherent state at very low temperatures. Assuming the orders are not spatially segregated, this  places constrains on the possible pairs of order parameters which are likely not to destroy each other immediately.\cite{Park2001,Mazin2009} In fact, the spin singlet  IVCW order parameter and the spin singlet TSC order parameter anticommute with each other, therefore they are allowed to coexist in general (although they may compete for Fermi surface density of state on the level of energetics). In contrast, a triplet IVCW order will serve as a pair breaking order with respect to the singlet TSC order, such that it will rapidly destroy superconductivity. As with superconductivity, a Zeeman field may stabilize the spin triplet IVCW at reduced temperatures, which would be interesting to explore in future experiments. Similarly, the spin triplet IVCW and superconducting orders are mutually compatible. The common origin of superconductivity and IVCW order implies that their transition temperatures should scale together if interactions are enhanced. Both orders should also be experimentally testable.

Finally, we have considered in detail topologically ordered Mott insulators arising from freezing the charge fluctuations in the candidate superconducting states. We show that condensing double-vortices in the spin-singlet chiral TSC leads to a chiral valley-spin liquid state in the Mott phase, where the time-reversal symmetry is broken spontaneously. Thus the valley degeneracy is lifted in the Mott insulator, consistent with the two-fold Landau level degeneracy in the quantum oscillation experiment. The coexistence of such an insulator with the superconductivity is also natural, as the two phases only differ by chargeon condensation, which can form puddles in the presence of inhomogeneity.

Our study already reveals a plethora of orders and their interrelations on the basis of approximate symmetries as well as quantum interference effects. Undoubtedly, this just the scratches the surface of an even richer set of exciting phenomena made possible in this new experimental platform.

We notice that several related works appear around the same time. Ref.\,\onlinecite{Isobe2018} focuses on the nesting among hotspots at the van Hove energy where the (CDW$'$,SDW$'$) and (singlet SC, triplet SC) orders correspond to the IVCW order $I^\mu$ and the inter-valley SC order $\Delta^\mu$ in our notation. Ref.\,\onlinecite{Wu2018} points out that a fluctuating $\O(n)$ vector order (with $n>2$) is crucial in explaining the emergence of SC inside the insulating phase. 

\begin{acknowledgements}
We thank Leon Balents, Shiang Fang, Bert Halperin, Pablo Jarillo-Herrero,  Efthimios Kaxiras, Allan MacDonald and Cenke Xu for useful discussions. AV is grateful to Adrian Po, T. Senthil and Liujun Zou for an earlier collaboration on related topics and several discussions. This work was supported by a Simons Investigator Grant. 
\end{acknowledgements}

\bibliography{WeakCoupling}

\appendix
\onecolumngrid
\section{Fermion Bilinear Operators and Local Interactions}\label{sec:AppA}
On a single orbital, the electron carries valley and spin degrees of freedom and can be written in the Majorana basis as
\begin{equation}
\chi=\mat{K\\ K'}\otimes\mat{\uparrow\\ \downarrow}\otimes\mat{\Re c\\ \Im c}.
\end{equation}
The Majorana basis spans an $8$-dimensional single-particle Hilbert space, in which there are altogether 28 fermion bilinear operators (as there are only 28 antisymmetric $8\times8$ matrices), which can be generally expressed as $\frac{1}{2}\chi^\intercal \sigma^{abc}\chi$ in terms of the Pauli operator $\sigma^{abc}\equiv\sigma^{a}\otimes\sigma^{b}\otimes\sigma^{c}$ (for $a,b,c=0,1,2,3$) with the constraint that $(\sigma^{abc})^\intercal=-\sigma^{abc}$. These operators can be classified according to their representations under the symmetry group $\U(4)=\U(1)_c\times\SU(4)$ or $\U(1)_c\times\U(1)_v\times\SO(4)$, as summarized in \tabref{tab:full}.
\begin{table}[htbp]
\caption{The number indicates the dimension of the representation (not irreducible for $\U(1)$ groups). The subscript labels the representation: $\U(1)$ group representations are labeled by their quantum numbers $q=0$ or $q=2$, non-Abelian group representations are labeled by names (sc - scalar, ve - vector, ad - adjoint, pv - pseudo-vector, ps - pseudo-scalar).}
\begin{center}
\begin{tabular}{|c|c|c|c|c|c|l|}
\hline
$\U(1)_c$ & $\SU(4)$ & $\U(1)_v$ & $\SO(4)$ & \multicolumn{2}{c|}
{operator} & order parameter\\ \hline
\multirow{6}{*}{$16_{0}$} & $1_\text{sc}$ & $1_{0}$ & $1_\text{sc}$ & $n_c$ & $\sigma^{002}$ & charge density \\ \cline{2-7}
& \multirow{5}{*}{$15_\text{ad}$} & \multirow{3}{*}{$7_{0}$} & $1_\text{ps}$ & $n_v$ & $\sigma^{302}$ & valley density \\ \cline{4-7}
& & & \multirow{2}{*}{$6_\text{ad}$} & \multirow{2}{*}{$\vect{S}_K\pm\vect{S}_{K'}$} & $\sigma^{012},\sigma^{020},\sigma^{032}$ & spin (FM/AFM \\
& & & & & $\sigma^{312},\sigma^{320},\sigma^{332}$ & between valleys)\\ \cline{3-7}
& & \multirow{2}{*}{$8_{2}$} & $4_\text{ve}$ & $I_x^\mu$ & $\sigma^{102},\sigma^{210},\sigma^{222},\sigma^{230}$ & \multirow{2}{*}{IVC$\times$(charge, spin)}\\ \cline{4-6}
& & & $4_\text{pv}$ & $I_y^\mu$ & $-\sigma^{200},\sigma^{112},\sigma^{120},\sigma^{132}$ & \\ \cline{1-7}
\multirow{6}{*}{$12_{2}$} & \multirow{3}{*}{$6_\text{ve}$} & $4_{0}$ & $4_\text{ve}$ & $\Re \Delta^\mu$ & $-\sigma^{121},\sigma^{233},-\sigma^{201},-\sigma^{213}$ & inter-valley SC\\ \cline{3-7}
& & \multirow{2}{*}{$2_{2}$} & $1_\text{sc}$ & $\Re(\Delta_K+\Delta_{K'})$ & $-\sigma^{021}$ & \multirow{2}{*}{intra-valley SC}\\ \cline{4-6}
& & & $1_\text{ps}$ & $\Im(\Delta_K-\Delta_{K'})$  & $\sigma^{323}$ & \\ \cline{2-7}
& \multirow{3}{*}{$6_\text{pv}$} & $4_{0}$ & $4_\text{pv}$ & $\Im \Delta^\mu$ & $\sigma^{123},\sigma^{231},\sigma^{203},-\sigma^{211}$ & inter-valley SC\\ \cline{3-7}
& & \multirow{2}{*}{$2_{2}$} & $1_\text{sc}$ & $\Im(\Delta_K+\Delta_{K'})$ & $\sigma^{023}$ & \multirow{2}{*}{intra-valley SC}\\ \cline{4-6}
& & & $1_\text{ps}$ & $\Re(\Delta_K-\Delta_{K'})$ & $-\sigma^{321}$ & \\ \hline
\end{tabular}
\end{center}
\label{tab:full}
\end{table}

One may therefore expect that the most generic $\U(1)_c\times\U(1)_v\times\SO(4)$ local interaction to be a linear combination of $n_c^2$, $n_v^2$, $\vect{S}_{K}^2+\vect{S}_{K'}^2$, $I^{\mu\dagger} I^\mu$ (with $I^\mu=I_x^\mu+\ii I_y^\mu$), $\Delta^{\mu\dagger}\Delta^\mu$, $\Delta_{K}^\dagger\Delta_{K}+\Delta_{K'}^\dagger\Delta_{K'}$ (exhausting all Fermion bilinear channels). However, these interaction terms are not linearly independent, as can be seen from
\begin{equation}
\begin{split}
&\frac{1}{4}(n_c^2+n_v^2)=-\frac{1}{6}(\vect{S}_{K}^2+\vect{S}_{K'}^2)=\frac{1}{4}(\Delta_{K}^\dagger\Delta_{K}+\Delta_{K'}^\dagger\Delta_{K'})=n_{K\uparrow}n_{K\downarrow}+n_{K'\uparrow}n_{K'\downarrow},\\
&\frac{1}{4}(n_c^2-n_v^2)=-\frac{1}{8}I^{\mu\dagger} I^\mu=\frac{1}{8}\Delta^{\mu\dagger}\Delta^\mu=(n_{K\uparrow}+n_{K\downarrow})(n_{K'\uparrow}+n_{K'\downarrow}).
\end{split}
\end{equation}
There are only two linearly independent local interactions, so the most generic local interaction should be
\begin{equation}
H_\text{int}=U_0(n_{K\uparrow}+n_{K\downarrow})(n_{K'\uparrow}+n_{K'\downarrow})+U_1(n_{K\uparrow}n_{K\downarrow}+n_{K'\uparrow}n_{K'\downarrow}).
\end{equation}

\section{Landau-Ginzburg Theory}\label{sec:LG}
In this appendix, we review the derivation of Landau-Ginzburg theory in Ref.\,\onlinecite{Xu2018} and propose a generalization beyond the mean-field framework. Our starting point is the BCS mean-field theory, described by the BdG Hamiltonian
\begin{equation}
\begin{split}
H_\text{BdG}&=\sum_{\vect{k}}\psi_\vect{k}^\dagger h_\text{BdG} \psi_\vect{k},\\
h_\text{BdG}&=\mat{\epsilon_\vect{k}&\Delta_\vect{k}^\dagger\\
\Delta_\vect{k}&-\epsilon_{\vect{k}}},
 \psi_\vect{k}=\mat{c_{K\vect{k}}\\
\ii\sigma^2 c^\dagger_{K'-\vect{k}}},
\end{split}
\end{equation}
where $c_{K\vect{k}}=(c_{K\vect{k}\uparrow},c_{K\vect{k}\downarrow})^\intercal$ describes the electrons around the $K$ valley and similarly for $c_{K'\vect{k}}$. The elements $\epsilon_\vect{k}$ and $\Delta_\vect{k}$ in $h_\text{BdG}$ are themself $2\times2$ matrices. We take the single band model proposed in the main text: $\epsilon_\vect{k}=(\vect{k}^2-\mu+\alpha(k_x^2-3k_xk_y^2))\sigma^0$. We consider the inter-valley pairing term $\Delta_\vect{k}$, which can be decomposed in $\O(4)$ components $\Delta_\vect{k}^\mu$, and each component is a linear combination of the leading form factors $w_\vect{k}$ and $w_\vect{k}^*$,
\begin{equation}\label{eq:Delta2}
\Delta_\vect{k}=\Delta_{\vect{k}}^\mu s^\mu=w_\vect{k} u^\mu s^\mu+w_\vect{k}^* v^\mu s^\mu,
\end{equation}
where $(s^0,\vect{s})=(\sigma^0,-\ii\vect{\sigma})$ and $u^\mu,v^\mu$ form complex four-component vectors. Since $\epsilon_\vect{k}$ is proportional to an identity matrix in the spin space, so the pairing gap is purely determined by the singular value of $\Delta_\vect{k}$. As $\Delta_\vect{k}$ is a $2\times2$ matrix, it will have two singular values, denoted as $\delta_{n\vect{k}}$ ($n=1,2$), corresponding to the gap for two different spin components (along certain direction determined by the singular vectors). These two singular values must be optimized independently, so the Landau-Ginzburg free energy should take the following form (to the quartic order of $\delta_{n\vect{k}}$)
\begin{equation}\label{eq:FLG1}
F=\sum_{n=1,2}\sum_{\vect{k}}r \delta_{n\vect{k}}^2+\kappa \delta_{n\vect{k}}^4+\cdots=\sum_\vect{k} r \Tr \Delta_\vect{k}^\dagger\Delta_\vect{k}+\kappa\Tr\Delta_\vect{k}^\dagger\Delta_\vect{k}\Delta_\vect{k}^\dagger\Delta_\vect{k}+\cdots.
\end{equation}
Plugging in \eqnref{eq:Delta2}, the above free energy reduces to
\begin{equation}\label{eq:FLG2}
\begin{split}
F&=\sum_\vect{k}r\Delta_\vect{k}^{\mu*}\Delta_\vect{k}^\mu+\kappa (2(\Delta_\vect{k}^{\mu*}\Delta_\vect{k}^\mu)^2-|\Delta_\vect{k}^\mu\Delta_\vect{k}^\mu|^2)+\cdots\\
&=\tilde{r}(u^{\mu*}u^\mu+v^{\mu*}v^\mu)+\tilde{\kappa}(2(u^{\mu*}u^\mu+v^{\mu*}v^\mu)^2+4|u^{\mu*}v^\mu|^2-4|u^{\mu}v^\mu|^2-|u^{\mu}u^\mu|^2-|v^{\mu}v^\mu|^2)+\cdots
\end{split}
\end{equation}
where the effective parameters are give by $\tilde{r}=r\sum_{\vect{k}\in\text{FS}}w_\vect{k}^*w_\vect{k}$ and $\tilde{\kappa}=\kappa\sum_{\vect{k}\in\text{FS}}(w_\vect{k}^*w_\vect{k})^2$. No matter what is the particular choice of the pairing form factor $w_\vect{k}$, $\tilde{r}$ and $\tilde{\kappa}$ always keep the same sign as $r$ and $\kappa$. The SC phase will correspond to the parameter regime of $\tilde{r}<0$ and $\tilde{\kappa}>0$. Within this parameter regime, we can minimize the free energy $F$ with respect to $u^\mu$, $v^\mu$. Only two distinct class of solutions are found: the chiral solution and the helical solution. They are degenerated in the free energy.

It turns out the above mean-field framework can not provide a solution for nematic superconductivity. Within the mean-field approach, as can be seen from \eqnref{eq:FLG2}, the Landau-Ginzburg free energy is simply a sum of the contributions from the order parameter $\Delta_\vect{k}$ at each momentum $\vect{k}$ independently. There is no interaction between the order parameters at different momenta. However such interaction in general could exist, which can be model by a more general free energy of the following form
\begin{equation}\label{eq:FLG3}
\begin{split}
F&=\sum_\vect{k}r\Delta_\vect{k}^{\mu*}\Delta_\vect{k}^\mu+\sum_{\vect{k},\vect{k}'}\kappa_{\vect{k}\vect{k}'}(2\Delta_{\vect{k}}^{\mu*}\Delta_{\vect{k}}^{\mu}\Delta_{\vect{k}'}^{\nu*}\Delta_{\vect{k}'}^{\nu}-\Delta_{\vect{k}}^{\mu*}\Delta_{\vect{k}}^{\mu*}\Delta_{\vect{k}'}^{\nu}\Delta_{\vect{k}'}^{\nu})+\cdots\\
&=\tilde{r}(u^{\mu*}u^\mu+v^{\mu*}v^\mu)+2\tilde{\kappa}_1(u^{\mu*}u^\mu+v^{\mu*}v^\mu)^2+4\tilde{\kappa}_2|u^{\mu*}v^\mu|^2-2(\tilde{\kappa}_1+\tilde{\kappa}_2)|u^{\mu}v^\mu|^2-\tilde{\kappa}_1(|u^{\mu}u^\mu|^2+|v^{\mu}v^\mu|^2)+\cdots.
\end{split}
\end{equation}
The momentum dependent coupling $\kappa_{\vect{k}\vect{k}'}$ describes the residual interaction between Cooper pairs, which would correspond to some eight-fermion interactions. Such interactions were not explicitly given in the model Hamiltonian, but they could definitely be generated under renormalization. In the case of $\kappa_{\vect{k}\vect{k}'}=\kappa\delta_{\vect{k}\vect{k}'}$, the free energy model \eqnref{eq:FLG3} reduces to \eqnref{eq:FLG2}. The generalized model in \eqnref{eq:FLG3} has more parameters to tune: $\tilde{\kappa}_1=\sum_{\vect{k},\vect{k}'\in\text{FS}}\kappa_{\vect{k}\vect{k}'}w_{\vect{k}}^*w_{\vect{k}}w_{\vect{k}'}^*w_{\vect{k}'}$ and $\tilde{\kappa}_2=\sum_{\vect{k},\vect{k}'\in\text{FS}}\kappa_{\vect{k}\vect{k}'}(w_{\vect{k}}^*w_{\vect{k}'})^2$. It is found that as long as $\tilde{\kappa}_2<0$, the nematic solution always minimizes the free energy. The solution is given in the main text.

\begin{figure}[htbp]
\begin{center}
\includegraphics[width=0.6\textwidth]{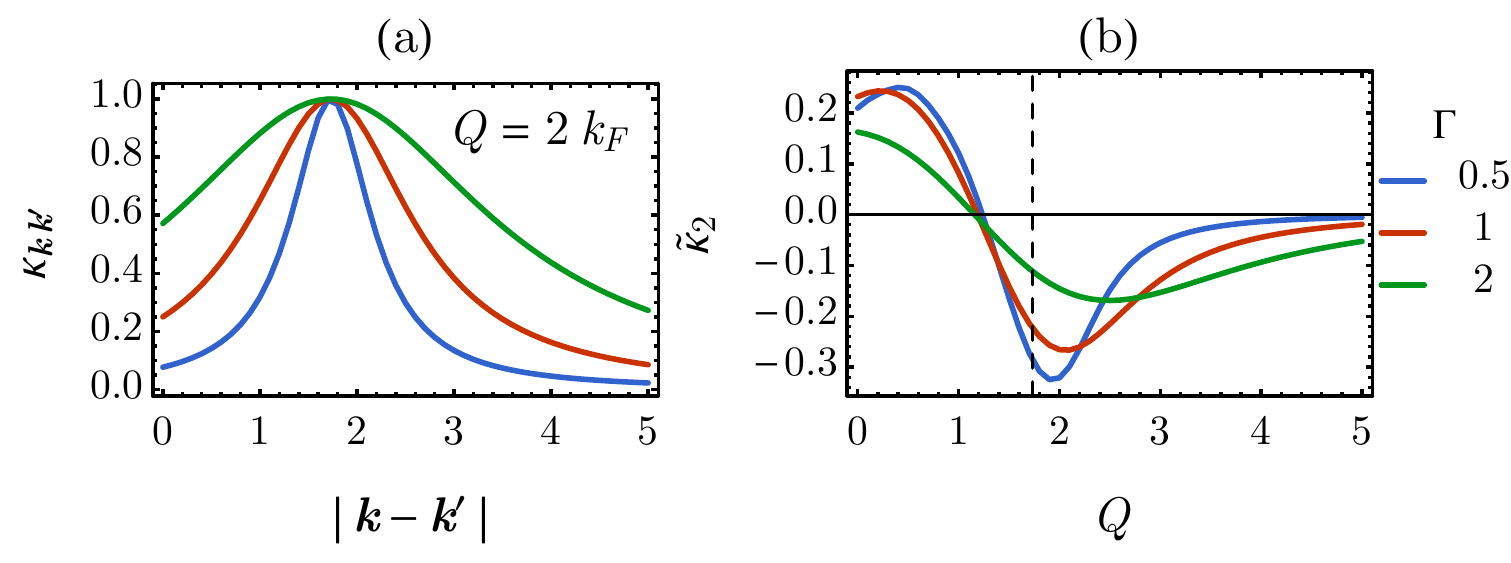}
\caption{(a) Dependence of $\kappa_{\vect{k}\vect{k}'}$ on $|\vect{k}-\vect{k}'|$ for different $\Gamma$. (b) $\tilde{\kappa}_2$ v.s. $Q$ for different $\Gamma$.}
\label{fig:kappa}
\end{center}
\end{figure}

To understand the possible scenario in which the nematic superconductivity might be favored, let us consider the following model of $\kappa_{\vect{k}\vect{k}'}$
\begin{equation}\label{eq:kappa}
\kappa_{\vect{k}\vect{k}'}=\frac{\kappa \Gamma^2}{(|\vect{k}-\vect{k}'|-Q)^2+\Gamma^2}.
\end{equation}
We assume that the interaction between Cooper pairs could depend on the difference between their relative (not center-of-mass) momenta $\vect{k}-\vect{k}'$. Suppose $\kappa_{\vect{k}\vect{k}'}$ has a non-trivial dependence on $\vect{k}-\vect{k}'$ which peaks around $|\vect{k}-\vect{k}'|\simeq Q$ with the width of the peak controlled by a parameter $\Gamma$, see \figref{fig:kappa}(a) for examples. An argument for such non-trivial momentum dependence is to note that when $Q\sim 2k_F$ (where $k_F\sim\sqrt{3\mu}$ is a crude notion of Fermi momentum even if the Fermi surface is not of circular shape), the Cooper pairs $\Delta_{k_F}^\mu$ and $\Delta_{-k_F}^\mu$ coincide and repel each other strongly, leading to a resonance of the repulsive interaction around $Q\sim 2k_F$. To ensure the stability of the Landau-Ginzburg theory, the quartic term would better be positive, i.e. $\kappa_{\vect{k}\vect{k}'}>0$. In terms of the model in \eqnref{eq:kappa}, it corresponds to setting $\kappa>0$. The precise form of $\kappa_{\vect{k}\vect{k}'}$ relies on many details, but the simple toy model in \eqnref{eq:kappa} already serves the purpose to show an important fact: it is possible to gain attractive interaction $\tilde{\kappa}_2<0$ even if $\kappa_{\vect{k}\vect{k}'}>0$ is always repulsive, as long as $\kappa_{\vect{k}\vect{k}'}$ has sufficiently non-trivial momentum dependence. Recall that $\tilde{\kappa}_2<0$ is all we need to stabilize the nematic superconductivity in the free energy analysis. To show this, we take the generic form factor $w_\vect{k}=w_d k_+^2+w_p k_-$ for a mixed $d+\ii d$ and $p-\ii p$ pairing, and calculate the coefficient $\tilde{\kappa}_2$ for different values of $Q$ and $\Gamma$. The result is shown in \figref{fig:kappa}(b). It demonstrates that around $Q\sim 2 k_F$, $\tilde{\kappa}_2$ indeed becomes negative. However this behavior heavily relies on the form of $\kappa_{\vect{k}\vect{k}'}$: if $\kappa_{\vect{k}\vect{k}'}$ does not peak around a sufficiently large momentum $Q$ (for example $Q\to0$), then $\tilde{\kappa}_2$ will be positive and not favoring the nematic superconductivity.

\section{Self-Consistent Mean-Field Theory}\label{sec:SCMF}
\begin{figure}[htbp]
\begin{center}
\includegraphics[width=0.48\columnwidth]{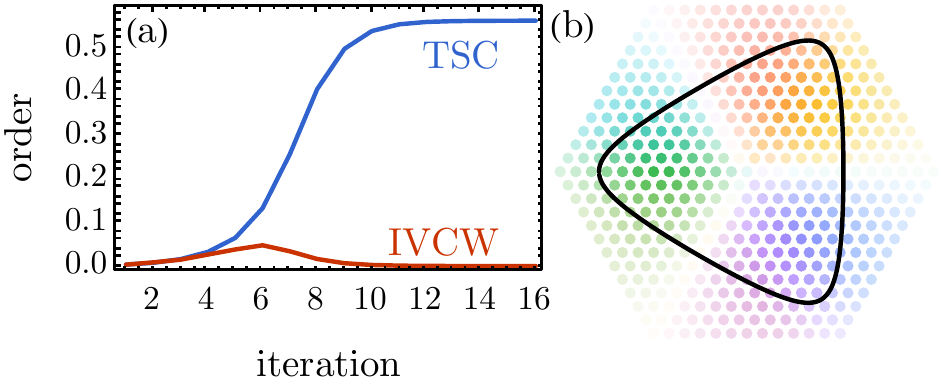}
\caption{(a) Self-consistent mean-field iteration showing the development and the convergence of the IVCW and the TSC order parameters. (b) The obtained gap function $\Delta_{\vect{k}}^0$ in the momentum space (around the $K$ pocket). Each dot represent a momentum point $\vect{k}$ on the momentum grid. The color indicates the phase of $\Delta_{\vect{k}}^0$, following the color scheme of \figref{fig:pairing}. The black curve marks out the shape of the $K$ pocket.}
\label{fig:SCMF}
\end{center}
\end{figure}

We provide here the details about the self-consistent mean-field theory to capture the competition between the IVCW and the TSC orders. Our starting point is the mean-field Hamiltonian $H_\text{MF}$ in the main text, which can be written in a more convenient form by introducing the fermion $\psi_{\vect{k}}=(c_{K\vect{k}\uparrow}, c_{K'-\vect{k}\downarrow}^{\dagger},c_{K' \vect{k}-\vect{Q}\uparrow},c_{K-\vect{k}+\vect{Q}\downarrow}^{\dagger})^\intercal$, such that
\begin{equation}\label{eq:MFdetails}
\begin{split}
H_\text{MF}&=\sum_{\vect{k}}\psi_{\vect{k}}^\dagger h_{\vect{k}}\psi_{\vect{k}}+H_\text{bg},\\
H_\text{bg}&=+g_II_\vect{Q}^{0*}I_\vect{Q}^0-g_\Delta\sum_\vect{k}\Delta_{-\vect{k}+\vect{Q}}^{0*}\Delta_{\vect{k}}^0,\\
h_{\vect{k}}&=\mat{
\epsilon_\vect{k} & -g_\Delta\Delta_\vect{k}^0 & g_II_\vect{Q}^{0*} & 0\\
-g_\Delta\Delta_\vect{k}^{0*} & -\epsilon_\vect{k} & 0 & -g_II_\vect{Q}^{0*}\\
g_II_\vect{Q}^{0} & 0 & \epsilon_{-\vect{k}+\vect{Q}} & -g_\Delta\Delta_{-\vect{k}+\vect{Q}}^0\\
0 & -g_II_\vect{Q}^{0} & -g_\Delta\Delta_{\vect{k}+\vect{Q}}^{0*} & -\epsilon_{-\vect{k}+\vect{Q}}}.
\end{split}
\end{equation}
We will take single-orbital model $\epsilon=\vect{k}^2-\mu+\alpha \Re k_+^3$ with $\alpha=1/3$. $I_\vect{Q}^0$ and $\Delta_{\vect{k}}^0$  are order parameters subject to optimization, where $I_\vect{Q}^0$ is treated as a constant independent of the momentum $\vect{k}$ while $\Delta_{\vect{k}}^0$ is a function of $\vect{k}$. To proceed, we focus on a patch of the momentum space that is large enough to cover the Fermi pockets, and we discretize the momentum patch to a triangular grid (to preserve the $C_3$ rotation symmetry) of $3L^2$ momentum points with $L=12$, see \figref{fig:SCMF}(b) for the illustration of momentum grid. We diagonalize the mean-field Hamiltonian at each momentum $\vect{k}$ to obtain the eigen energy $E_{n\vect{k}}$, s.t. $h_{\vect{k}}\ket{n}=E_{n\vect{k}}\ket{n}$. The free energy of the fermion can be evaluated from $E_{n\vect{k}}$ as
\begin{equation}
F_\text{MF}=-\beta^{-1}\sum_{n\vect{k}}\ln(1+e^{-\beta E_{n\vect{k}}})+H_\text{bg}.
\end{equation}
The free energy $F_\text{MF}$ is a function of the order parameters $I_\vect{Q}^0,\Delta_\vect{k}^0$. Given the temperature $T$ and the chemical potential $\mu$, we can find the optimal configuration of the order parameters that minimize $F_\text{MF}$. For example, \figref{fig:SCMF} shows one typical result of the mean-field iteration in the TSC phase. The $d+\ii d$ and $p-\ii p$ mixed paring form factor is clearly seen from \figref{fig:SCMF}(b), consistent with the result in \figref{fig:pairing}(b) obtained by a different method (by solving the gap equation). Repeating this calculation, we can find the mean-field solution numerically throughout the phase diagram.

\begin{figure}[htbp]
\begin{center}
\includegraphics[width=0.3\columnwidth]{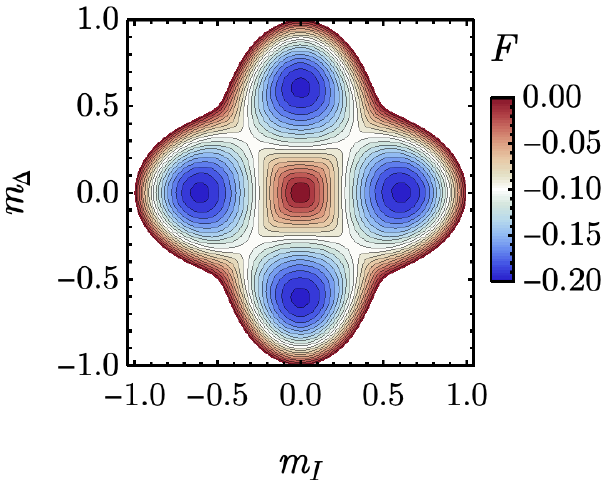}
\caption{Contour plot of free energy $F$ in \eqnref{eq:free energy} as a function of the gaps $m_I$ and $m_\Delta$. The IVCW and the TSC gaps repel each other, so they can not coexist.}
\label{fig:free energy}
\end{center}
\end{figure}

It is found that the IVCW order $I_\vect{Q}^0$ and the TSC order $\Delta_{\vect{k}}^0$ never coexist in the phase diagram. In fact, they are not expected to coexist, because they are commuting order parameters that always compete for Fermi surface density of state. This mechanism can be understood from a simplified toy model. In the case of perfect nesting, we have $\Delta_\vect{k}^0=-\Delta_{-\vect{k}+\vect{Q}}^0$ and $\epsilon_\vect{k}=-\epsilon_{-\vect{k}+\vect{Q}}$. Let us ignore the momentum dependence along the Fermi surface and simply set $\epsilon_\vect{k}=k_\perp$ ($k_\perp$ is the momentum perpendicular to the Fermi surface), $-g_\Delta\Delta_\vect{k}^0=m_\Delta$, $g_I I_\vect{Q}^0=m_I$, then the mean-field single-particle Hamiltonian $h_\vect{k}$ in \eqnref{eq:MFdetails} takes the form of
\begin{equation}
h=\mat{
k_\perp & m_\Delta & m_I & 0\\
m_\Delta & -k_\perp & 0 & -m_I\\
m_I & 0 & -k_\perp & -m_\Delta\\
0 & -m_I & -m_\Delta & k_\perp}=k_\perp \sigma^{33}+m_I \sigma^{13}+m_\Delta \sigma^{31}.
\end{equation}
The notation $\sigma^{ab}=\sigma^a\otimes\sigma^b$ stands for the tensor product of the Pauli matrices. $m_I$ and $m_\Delta$ are promotional to the IVCW and the TSC gaps respectively. By saying that the two orders commute, we mean their corresponding vertex matrices $\sigma^{13}$ and $\sigma^{31}$ commute. As a result of the commutativity between IVCW and TSC orders, the eigenenergies of $h$ are $\pm\sqrt{k_\perp^2+(m_I\pm m_\Delta)^2}$, which give rise to two gaps $|m_I\pm m_\Delta|$ for the Fermi surfaces in general. However, it is energetically favorable to gap out all Fermi surfaces with the same gap size. This can be seen by evaluating the free energy at zero temperature,
\begin{equation}\label{eq:free energy}
\begin{split}
F(m_I,m_\Delta)&\propto-\int_{-\Lambda}^{\Lambda}\dd k_\perp\Big(\sqrt{k_\perp^2+(m_I+m_\Delta)^2}+\sqrt{k_\perp^2+(m_I-m_\Delta)^2}\Big)\\
&\simeq-\frac{1}{4}(m_I+m_\Delta)^2 \ln\frac{\Lambda^2}{(m_I+m_\Delta)^2}-\frac{1}{4}(m_I-m_\Delta)^2 \ln\frac{\Lambda^2}{(m_I-m_\Delta)^2}.
\end{split}
\end{equation}
With a choice of the momentum cutoff at $\Lambda=1$, the free energy profile is shown in \figref{fig:free energy}. One can see that the gaps (order parameters) $m_I$ and $m_\Delta$ repel each other and the free energy is minimized only if one of them vanishes, i.e. $m_I=0$ or $m_\Delta=0$. Therefore the two orders are not expected to coexist.

In the self-consistent mean-field iteration of the full model in \eqnref{eq:MFdetails}, we also explicitly tested the possibility of coexisting IVCW and TSC orders. We start with the initial condition that both order parameter are non-zero and providing roughly the same size of the gap. We found that under the self-consistent iteration (equivalent to the gradient decent to minimize the free energy), the oder parameter always flow to either the IVCW or the TSC fixed point (with one kind of order only), see \figref{fig:SCMF}(a) for example. We performed this test for several different choices of the model parameters near the IVCW-TSC transition. We did not observe a mean-field saddle point with coexisting orders.

\end{document}